%% file: ton_airsync.tex
\documentclass[journal, twocolumn]{IEEEtran}

\usepackage{amsmath}
\usepackage{amssymb}
\usepackage{amsfonts}
\usepackage{graphicx}

\usepackage{cite}
\usepackage{subfigure}
\usepackage{url}
\usepackage{array}
\usepackage{verbatim}
\usepackage{widetext}
\usepackage{comment}
\usepackage{stfloats}

\usepackage{color}
\definecolor{red}{rgb}{1,0,0}
\definecolor{blue}{rgb}{0,0,1}
\newcommand{\kostas}[1]{{}}
\newcommand{\vlad}[1]{{}}

\newcounter{MYtempeqncnt}

\begin{document}
\title{AirSync: Enabling Distributed Multiuser MIMO with Full Spatial Multiplexing}

\author{Horia Vlad Balan, Ryan Rogalin, Antonios Michaloliakos, Konstantinos Psounis,~\IEEEmembership{Senior Member, IEEE} and 
Giuseppe Caire,~\IEEEmembership{Fellow, IEEE}%
\IEEEcompsocitemizethanks{
\IEEEcompsocthanksitem All authors are with the University of Southern California.
}

\thanks{This work was partially funded by a grant of the Ming-Hsieh Institute, University of Southern California.}}

\maketitle

\input{abstract}

\input{macros}

\input{introduction}

\input{relatedwork}

\input{idea}

\input{system}

\input{performance}
\input{medium}

\input{discussion}

\bibliographystyle{abbrv}
\bibliography{IEEEabrv,paper}

\end{document}

%% file: abstract.tex
\begin{abstract}

The enormous success of advanced wireless devices is pushing the
demand for higher wireless data rates. Denser spectrum reuse through
the deployment of more Access Points (APs) per square mile has the potential
to successfully meet the increasing demand for more bandwidth. 
In principle, distributed multiuser MIMO (MU-MIMO) provides the best
approach to infrastructure density increase, since several access points are connected to a central server 
and operate as a large distributed multi-antenna access point.  This ensures that all
transmitted signal power serves the purpose of data transmission,
rather than creating interference.
In practice, however, a number of implementation difficulties must be addressed, 
the most significant of which is aligning the
phases of all jointly coordinated APs.

In this paper we propose AirSync, a novel scheme which provides 
timing and phase synchronization accurate enough to enable 
distributed MU-MIMO. 
AirSync detects the slot boundary such that all APs are time-synchronous within a cyclic prefix (CP) 
of the  OFDM modulation, and predicts the instantaneous carrier phase correction along the transmit slot
such that all transmitters maintain their coherence, which is necessary for multiuser beamforming. 
We have implemented AirSync as a digital circuit in the FPGA of the
WARP radio platform. Our experimental testbed, comprised of four
APs and four clients, shows that AirSync is able to achieve timing synchronization within the OFDM CP and 
carrier phase coherence (after the correction) 
within a few degrees. For the purpose of demonstration, we have implemented 
two MU-MIMO precoding schemes, Zero-Forcing Beamforming (ZFBF) and
Tomlinson-Harashima Precoding (THP). In both cases our system approaches
the theoretical optimal multiplexing gains. We also discuss aspects related to the MAC and multiuser 
scheduling design, in relation to the distributed MU-MIMO architecture. 
To the best of our knowledge, AirSync offers the first ever realization of the full
distributed MU-MIMO multiplexing gain, namely the ability to increase the number of
active wireless clients per time-frequency slot linearly with the number of jointly coordinated
APs, without reducing the per client rate.

\end{abstract}

%% file: macros.tex
\setlength\unitlength{1mm}

\newcommand{\insertfig}[3]{
\begin{figure}[htbp]\begin{center}\begin{picture}(120,90)
\put(0,-5){\includegraphics[width=12cm,height=9cm,clip=]{#1.eps}}\end{picture}\end{center}
\caption{#2}\label{#3}\end{figure}}

\newcommand{\insertxfig}[4]{
\begin{figure}[htbp]
\begin{center}
\leavevmode \centerline{\resizebox{#4\textwidth}{!}{\input
#1.pstex_t}}
%\vspace*{-0.2in}
\caption{#2} \label{#3}
\end{center}
\end{figure}}

\long\def\comment#1{}

% bb font symbols

\newfont{\bbb}{msbm10 scaled 700}
\newcommand{\CCC}{\mbox{\bbb C}}

\newfont{\bb}{msbm10 scaled 1100}
\newcommand{\CC}{\mbox{\bb C}}
\newcommand{\PP}{\mbox{\bb P}}
\newcommand{\RR}{\mbox{\bb R}}
\newcommand{\QQ}{\mbox{\bb Q}}
\newcommand{\ZZ}{\mbox{\bb Z}}
\newcommand{\FF}{\mbox{\bb F}}
\newcommand{\GG}{\mbox{\bb G}}
\newcommand{\EE}{\mbox{\bb E}}
\newcommand{\NN}{\mbox{\bb N}}
\newcommand{\KK}{\mbox{\bb K}}

% Vectors

\newcommand{\av}{{\bf a}}
\newcommand{\bv}{{\bf b}}
\newcommand{\cv}{{\bf c}}
\newcommand{\dv}{{\bf d}}
\newcommand{\ev}{{\bf e}}
\newcommand{\fv}{{\bf f}}
\newcommand{\gv}{{\bf g}}
\newcommand{\hv}{{\bf h}}
\newcommand{\iv}{{\bf i}}
\newcommand{\jv}{{\bf j}}
\newcommand{\kv}{{\bf k}}
\newcommand{\lv}{{\bf l}}
\newcommand{\mv}{{\bf m}}
\newcommand{\nv}{{\bf n}}
\newcommand{\ov}{{\bf o}}
\newcommand{\pv}{{\bf p}}
\newcommand{\qv}{{\bf q}}
\newcommand{\rv}{{\bf r}}
\newcommand{\sv}{{\bf s}}
\newcommand{\tv}{{\bf t}}
\newcommand{\uv}{{\bf u}}
\newcommand{\wv}{{\bf w}}
\newcommand{\vv}{{\bf v}}
\newcommand{\xv}{{\bf x}}
\newcommand{\yv}{{\bf y}}
\newcommand{\zv}{{\bf z}}
\newcommand{\zerov}{{\bf 0}}
\newcommand{\onev}{{\bf 1}}

% Matrices

\newcommand{\Am}{{\bf A}}
\newcommand{\Bm}{{\bf B}}
\newcommand{\Cm}{{\bf C}}
\newcommand{\Dm}{{\bf D}}
\newcommand{\Em}{{\bf E}}
\newcommand{\Fm}{{\bf F}}
\newcommand{\Gm}{{\bf G}}
\newcommand{\Hm}{{\bf H}}
\newcommand{\Id}{{\bf I}}
\newcommand{\Jm}{{\bf J}}
\newcommand{\Km}{{\bf K}}
\newcommand{\Lm}{{\bf L}}
\newcommand{\Mm}{{\bf M}}
\newcommand{\Nm}{{\bf N}}
\newcommand{\Om}{{\bf O}}
\newcommand{\Pm}{{\bf P}}
\newcommand{\Qm}{{\bf Q}}
\newcommand{\Rm}{{\bf R}}
\newcommand{\Sm}{{\bf S}}
\newcommand{\Tm}{{\bf T}}
\newcommand{\Um}{{\bf U}}
\newcommand{\Wm}{{\bf W}}
\newcommand{\Vm}{{\bf V}}
\newcommand{\Xm}{{\bf X}}
\newcommand{\Ym}{{\bf Y}}
\newcommand{\Zm}{{\bf Z}}

% Calligraphic

\newcommand{\Ac}{{\cal A}}
\newcommand{\Bc}{{\cal B}}
\newcommand{\Cc}{{\cal C}}
\newcommand{\Dc}{{\cal D}}
\newcommand{\Ec}{{\cal E}}
\newcommand{\Fc}{{\cal F}}
\newcommand{\Gc}{{\cal G}}
\newcommand{\Hc}{{\cal H}}
\newcommand{\Ic}{{\cal I}}
\newcommand{\Jc}{{\cal J}}
\newcommand{\Kc}{{\cal K}}
\newcommand{\Lc}{{\cal L}}
\newcommand{\Mc}{{\cal M}}
\newcommand{\Nc}{{\cal N}}
\newcommand{\Oc}{{\cal O}}
\newcommand{\Pc}{{\cal P}}
\newcommand{\Qc}{{\cal Q}}
\newcommand{\Rc}{{\cal R}}
\newcommand{\Sc}{{\cal S}}
\newcommand{\Tc}{{\cal T}}
\newcommand{\Uc}{{\cal U}}
\newcommand{\Wc}{{\cal W}}
\newcommand{\Vc}{{\cal V}}
\newcommand{\Xc}{{\cal X}}
\newcommand{\Yc}{{\cal Y}}
\newcommand{\Zc}{{\cal Z}}

% Bold greek letters

\newcommand{\alphav}{\hbox{\boldmath$\alpha$}}
\newcommand{\betav}{\hbox{\boldmath$\beta$}}
\newcommand{\gammav}{\hbox{\boldmath$\gamma$}}
\newcommand{\deltav}{\hbox{\boldmath$\delta$}}
\newcommand{\etav}{\hbox{\boldmath$\eta$}}
\newcommand{\lambdav}{\hbox{\boldmath$\lambda$}}
\newcommand{\epsilonv}{\hbox{\boldmath$\epsilon$}}
\newcommand{\nuv}{\hbox{\boldmath$\nu$}}
\newcommand{\muv}{\hbox{\boldmath$\mu$}}
\newcommand{\zetav}{\hbox{\boldmath$\zeta$}}
\newcommand{\phiv}{\hbox{\boldmath$\phi$}}
\newcommand{\psiv}{\hbox{\boldmath$\psi$}}
\newcommand{\thetav}{\hbox{\boldmath$\theta$}}
\newcommand{\tauv}{\hbox{\boldmath$\tau$}}
\newcommand{\omegav}{\hbox{\boldmath$\omega$}}
\newcommand{\xiv}{\hbox{\boldmath$\xi$}}
\newcommand{\sigmav}{\hbox{\boldmath$\sigma$}}
\newcommand{\piv}{\hbox{\boldmath$\pi$}}
\newcommand{\rhov}{\hbox{\boldmath$\rho$}}

\newcommand{\Gammam}{\hbox{\boldmath$\Gamma$}}
\newcommand{\Lambdam}{\hbox{\boldmath$\Lambda$}}
\newcommand{\Deltam}{\hbox{\boldmath$\Delta$}}
\newcommand{\Sigmam}{\hbox{\boldmath$\Sigma$}}
\newcommand{\Phim}{\hbox{\boldmath$\Phi$}}
\newcommand{\Pim}{\hbox{\boldmath$\Pi$}}
\newcommand{\Psim}{\hbox{\boldmath$\Psi$}}
\newcommand{\Thetam}{\hbox{\boldmath$\Theta$}}
\newcommand{\Omegam}{\hbox{\boldmath$\Omega$}}
\newcommand{\Xim}{\hbox{\boldmath$\Xi$}}
\newcommand{\Csf}{{\sf C}}

% mixed symbols

\newcommand{\sinc}{{\hbox{sinc}}}
\newcommand{\diag}{{\hbox{diag}}}
\renewcommand{\det}{{\hbox{det}}}
\newcommand{\trace}{{\hbox{tr}}}
\newcommand{\sign}{{\hbox{sign}}}
\renewcommand{\arg}{{\hbox{arg}}}
\newcommand{\var}{{\hbox{var}}}
\newcommand{\cov}{{\hbox{cov}}}
\newcommand{\SINR}{{\sf sinr}}
\newcommand{\SNR}{{\sf snr}}
\newcommand{\Ei}{{\rm E}_{\rm i}}
\renewcommand{\Re}{{\rm Re}}
\renewcommand{\Im}{{\rm Im}}
\newcommand{\eqdef}{\stackrel{\Delta}{=}}
\newcommand{\defines}{{\,\,\stackrel{\scriptscriptstyle \bigtriangleup}{=}\,\,}}
\newcommand{\<}{\left\langle}
\renewcommand{\>}{\right\rangle}
\newcommand{\herm}{{\sf H}}
\newcommand{\trasp}{{\sf T}}
\newcommand{\transp}{{\sf T}}
\renewcommand{\vec}{{\rm vec}}

\newcommand{\RED}{\color[rgb]{1.00,0.10,0.10}}
\newcommand{\BLUE}{\color[rgb]{0,0,0.90}}
\newcommand{\GREEN}{\color[rgb]{0,0.80,0.20}}

%% file: introduction.tex
\section{Introduction}
\label{sec:introduction}

The enormous success of advanced wireless devices such as tablets and smartphones 
is pushing the demand for higher and higher wireless data rates and is causing significant stress 
to existing networks.
While new standards (e.g., 802.11n/ac and 4G) are developed almost every couple of years, 
novel and more radical approaches to this problem are yet to be tested. 
The fundamental bottleneck is that wireless bandwidth is simply upper bounded 
by physical laws, in contrast to wired bandwidth, where putting new fiber on the ground has been the de-facto 
solution for decades. 
Advances in network protocols, modulation and coding schemes
have managed steady but relatively modest spectral efficiency (bit/s/Hz) improvements.
4G-LTE, for instance, offers two to five times better spectral efficiency than 2.5G-EDGE.
Denser spectrum reuse, i.e., placing more access points per square mile, 
has the potential to successfully meet the increasing demand for more wireless bandwidth 
\cite{cisco_report}. 
On the other hand, in contrast to conventional cellular systems, 
a very dense infrastructure deployment cannot be carefully  planned and managed for reasons pertaining to scale
and cost. Therefore, the multiuser interference between different uncoordinated Access Points (APs) 
represents the main system bottleneck  to achieve truly high spectral efficiency. 

In theory, the ultimate answer to this problem is distributed
multiuser MIMO (also known as ``virtual MIMO''), where several
(possibly multi-antenna) APs are connected to central
server and operate as a large distributed multi-antenna base station.
When using joint decoding in the uplink and joint precoding in the
downlink, all transmitted signal power is useful, as opposed to
conventional random access scenarios (e.g., carrier-sense) which waste power through
interference. This approach is particularly
suited to the case of an enterprise network (e.g., a WLAN covering a
conference center, an airport terminal or a university), or to the case of
clusters of closely spaced home networks connected to the Internet
infrastructure through the same cable bundle.

Distributed multiuser MIMO (MU-MIMO) is regarded today mostly as a
theoretical solution because of some serious implementation hurdles,
such as providing accurate timing and carrier phase reference 
to all jointly coordinated APs and the ability to perform efficient joint precoding 
at a central server connected to the APs through a wired backhaul of limited capacity.

\begin{figure}[t]
 \centering
 \includegraphics[width=.44\textwidth]{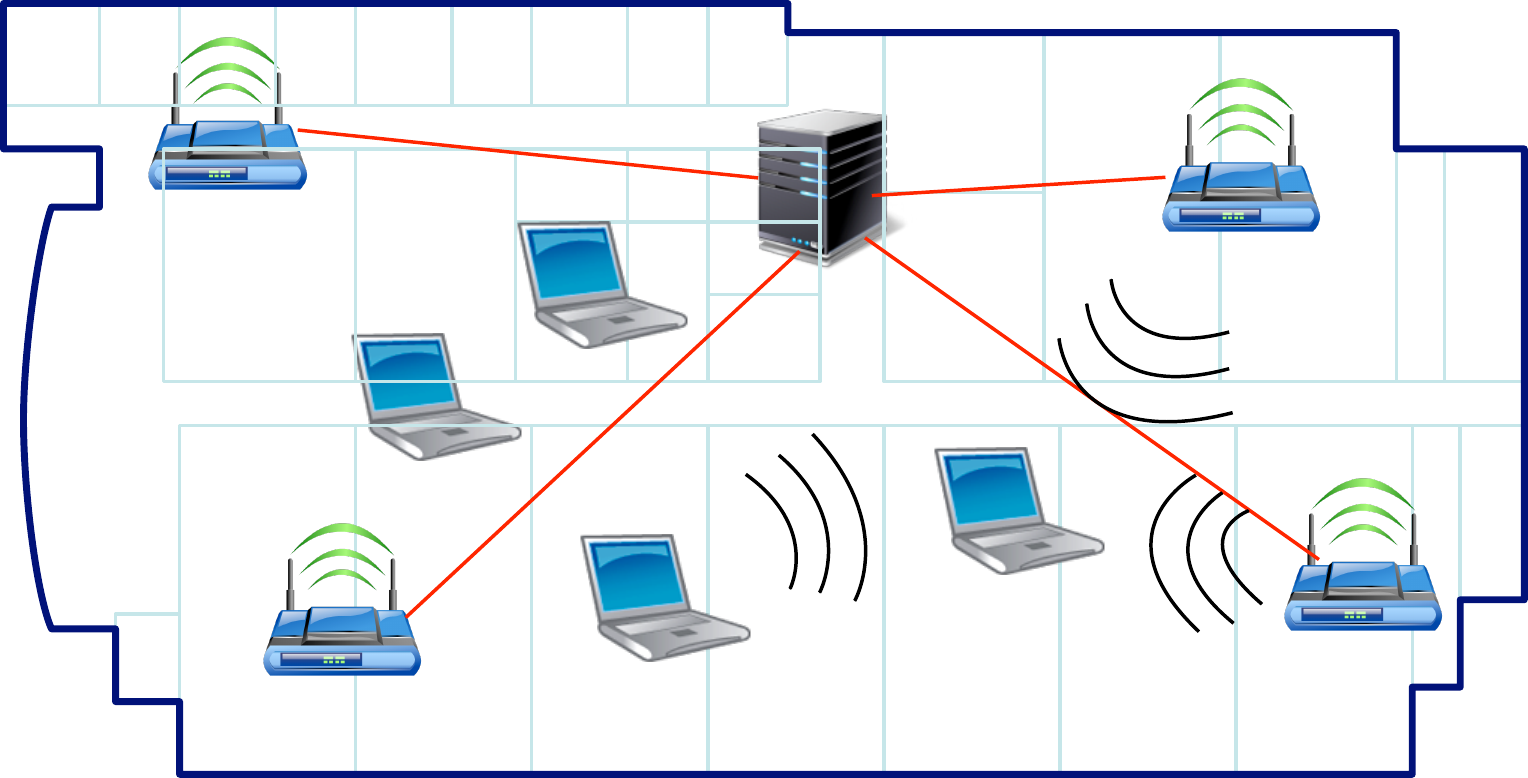}
 \caption{Enterprise Wifi and Distributed MIMO. \textnormal{Multiple
     access points connected to a central server through Ethernet (red
     lines) coordinate their transmissions to several clients by using
     distributed MIMO.}}
 \label{wifi}
\end{figure}

We consider a typical enterprise network as illustrated in Figure \ref{wifi}. Since in such networks the wired backhaul 
is fast enough
to allow for efficient joint processing at the server (see Section
\ref{sec:hardware}), the major obstacle to achieving the full
distributed MU-MIMO multiplexing gain is represented by the lack of
synchronization between the jointly processed APs.  The perceived
difficulty of this task has led some researchers to believe that it is
practically impossible to achieve full multiplexing gain in the
context of distributed MU-MIMO. In this paper, we present the first
(to the best of our knowledge) real-world testbed implementation which
achieves the theoretical optimal gain by correcting, in real time, the
instantaneous phase offsets between geographically separated access
points.  We achieve this via \emph{AirSync}.

In a nutshell, AirSync tracks the instantaneous phase of a pilot
signal broadcasted by a reference AP (the master AP), and predicts the
phase correction across the duration of the MU-MIMO precoded slot in
order to de-rotate the complex baseband signal samples. This enables APs to
maintain phase coherence, which is
necessary for MU-MIMO precoding. Notice that each AP {\em transmits}
the precoded data signal and {\em receives} the master AP pilot signal
simultaneously. This is accomplished by dedicating one antenna per AP
to pilot reception, while the others are used for MU-MIMO
transmission.  We have implemented AirSync as a digital circuit in the
FPGA of the WARP radio platform \cite{warp}.  We have also implemented
\emph{Zero-Forcing Beamforming} \cite{caire-shamai} and
\emph{Tomlinson-Harashima Precoding} \cite{windpassinger-huber}, two
popular MU-MIMO precoding schemes widely investigated, from a
theoretical viewpoint, in the literature. Using Airsync in a testbed
consisting of eight WARP radios, four acting as access points
connected to a central server and four acting as clients, we have
shown that the theoretical optimal gain of multiuser MIMO is
achievable in practice.

Optimal scheduling of downlink transmissions involves taking into
account channel state information. Consequently, we investigate the
protocol design for the MAC layer in distributed MU-MIMO systems,
including channel-based client selection, downlink scheduling and
adaptive coding modulation, either obtained through a family of
Modulation and Coding Schemes (MCSs), as in IEEE 802.11n
\cite{802.11ac}, or through the use of Incremental Redundancy rateless
coding at the physical layer \cite{raptor} \cite{spinal}.

Several extensions and improvements of this basic layout are possible,
and are discussed in the paper. For example, the master pilot signal
range can be extended by regenerating and repeating the pilot signal
at different frequencies. Also, we discuss the possibility of
estimating the downlink channel matrix from training signals in the
uplink and exploit the physical channel reciprocity of Time-Division
Duplexing (TDD). In particular, this latter issue is discussed in
detail in the recent work \cite{argos} for a large centralized
MU-MIMO system where all transmit antennas are clocked together (both
timing and carrier frequency) and therefore are perfectly synchronous.
Thanks to AirSync, the same approach for calibration can be used in a
distributed MU-MIMO system, although in the present implementation we
use a more conventional downlink training and feedback configuration.

While recently there have been a number of very interesting
and important works in which some of the gains of multiuser MIMO have
been shown (see Section \ref{sec:relatedwork} for more details) none
of these has managed to achieve both timing and carrier phase synchronization 
between remote transmitters precise enough to implement MU-MIMO with optimal multiplexing gain 
in the distributed transmitters scenario. It is also worthwhile to notice that while single-user MIMO and single AP (centralized) MU-MIMO
can offer multiplexing gains for a given configuration of transmit and receive antennas, these can be further 
increased by extending the cooperation to the distributed case, provided that the transmitters (the APs) 
can be synchronized with sufficient accuracy. Therefore, our approach can potentially provide {\em additional multiplexing gains}
on top of an existing configuration. 

In summary, in this paper we make the following contributions: 
\begin{itemize}
\item We introduce AirSync, the first (to the best of our knowledge) scheme which achieves phase synchronization in a distributed MU-MIMO setting.
\item We implement AirSync as a digital circuit in the FPGA of the WARP platform.
\item We showcase in a testbed consisting of eight WARP radios that, thanks to AirSync, the theoretically optimal 
spatial multiplexing gain is achievable in practice.
\item We discuss practical implementation aspects of the MAC layer for distributed MU-MIMO, including scheduling, user selection, 
and adaptive MCS versus incremental redundancy with rateless coding. 
\end{itemize}

We conclude this introduction by providing a brief outline for the
rest of the paper.  In Section \ref{sec:relatedwork} we discuss in
detail related work both on the theoretical side and on the practical
side. In Section \ref{sec:primer} we offer an overview of 
the MU-MIMO precoding techniques implemented in our work. 
In Section \ref{sec:idea} we show why phase synchronization is needed to achieve the
promised gain, and describe, in general terms, AirSync.  
In Section \ref{sec:hardware} we present the hardware implementation of AirSync
in detail.  In Section \ref{section:performance} we present a number
of results obtained using our testbed implementation with four access
points and four clients.  We show results regarding the
synchronization accuracy, the beamforming gain, the zero-forcing
precision and the implementation of ZFBF and THP as the MU-MIMO precoding schemes to achieve optimal 
multiuser multiplexing gains. Section \ref{section:medium}  discusses the issues related to
the MAC design for distributed MU-MIMO. Finally, Section
\ref{sec:discussion} points out some developments under on-going investigation.

%% file: relatedwork.tex
\section{Related Work}
\label{sec:relatedwork}

{\bf Theoretical Foundations.}
The pioneering papers by Foschini \cite{foschini} and Telatar
\cite{telatar_mimo_cap} have shown that adding multiple antennas both
to the transmitter and to the receiver increases the capacity of a
point-to-point communication channel. At practical medium-to-high
Signal to Noise Ratios (SNRs), this gain manifests as a multiplicative
factor equal to the rank of the matrix representing the transfer
function between the transmit and the receive antennas. For
sufficiently rich propagation scattering, with probability 1 this
factor is equal to $\min\{N_t,N_r\}$, where $N_t$ and $N_r$ denote the
number of transmit and receive antennas, respectively.  The MIMO
capacity gain can be interpreted as the implicit ability to create
$\min\{N_t,N_r\}$ ``parallel'' non-interfering channels corresponding
to the channel matrix eigenmodes, and it is referred to in the
literature as {\em multiplexing gain}, or as the {\em degrees of
  freedom} of the channel.  Subsequently, Caire and Shamai
\cite{caire-shamai} have shown that the MIMO broadcast channel, where
the transmitter has $N_t$ antennas and serves $K$ clients with $N_r$
antennas each, exhibits an analogous capacity factor increase of
$\min\{N_t, KN_r\}$, suggesting that a transmitter with multiple
antennas could transmit simultaneously on the same frequency to
independent users.  Such multiuser communication has two additional
requirements. First, precoding of the transmitted data is needed to
prevent the different spatial streams from mutually interfering.
Second, the transmitter requires accurate knowledge of the channel
matrix (channel state information) in order to realize this precoding.

The idea of precoding has spurred research beyond the scope of this
paper. Dirty Paper Coding (DPC) \cite{costa-dpc} with a Gaussian
coding ensemble achieves the capacity of the MIMO broadcast channel
\cite{weingarten}, but is difficult to implement in practice. The
well-known linear Zero-Forcing Beamforming (ZFBF) \cite{caire-shamai}
achieves the same high-SNR capacity factor increase, with some fixed
gap from optimal that can be reduced when the number of clients is
large and the transmitter can dynamically select the clients to be
served depending on their channel state information
\cite{yoo-goldsmith, caire_isit}.  Tomlinson-Harashima Precoding is
another well-studied, but infrequently implemented technique, which
efficiently approximates DPC at high SNRs \cite{windpassinger-huber}.
A number of other precoding strategies (e.g., lattice reduction,
regularized vector perturbation) have been studied and the interested
reader is referred to \cite{spencer2004introduction} and references
therein.  For the purposes of this paper ZFBF and THP will be the
primary methods of interest because of their conceptual simplicity and
good complexity/performance tradeoff.

{\bf Practical Implementations.}  A number of recent system
implementations have made forays into the topics of multiuser MIMO
transmission and distributed, slot aligned OFDM transmission. MU-MIMO
ZFBF as a precoding scheme in a centralized setting have been examined
in \cite{aryafar10design}, for a system consisting of a single AP with
multiple antennas hosted on the same radio board.  The use of
interference alignment and cancellation as a precoding technique,
which does not require slot synchronization 
or phase synchronization of the transmitters, has been
illustrated in \cite{gollakota09interference}. While this solution
achieves a part of the potential spatial multiplexing gain, in order
to realize the full spatial multiplexing requires tight phase
synchronization between the jointly precoded transmitters
\cite{lapidoth_allerton,vaze_degrees}.

In order to be able to adopt the classical discrete-time
symbol-synchronous complex baseband equivalent channel models used in
communication and information theory, the fundamental underlying
assumption is that transmissions from different nodes align within the
cyclic prefix of OFDM (referred to as ``slot alignment'' in the
following).  If this is not verified, then inter-block interference
arises and the channel does not decompose any longer into a set of
discrete-time parallel channels.  Slot alignment was used in
SourceSync \cite{rahul10sourcesync} in conjunction with space-time
block coding in order to provide a diversity gain in a distributed
MIMO downlink system. In Fine-Grained Channel Access \cite{tan10fine},
a similar technique allows for multiple independent clients to share
the frequency band in fine increments, without a need for guard bands,
resulting in a flexible OFDMA (OFDM with orthogonal multiple access)
uplink implementation.

Distributed space-time coding and flexible orthogonal access do not
increase the system degrees of freedom, since at most a single
information symbol per time-frequency dimension can be transmitted.
\footnote{A time-frequency dimension escorresponds to one symbol in the
  frequency domain, spanning one OFDM subcarrier over one OFDM symbol
  duration, and spans (approximately) 1 s$\times$Hz.}

%% file: idea.tex
\section{A Multiuser MIMO Primer}
\label{sec:primer}

We consider the OFDM signaling format, 
as in the last generation of WLANs and cellular systems (e.g.,
IEEE 802.11a/g/n and 4G-LTE \cite{molischbook}). 
OFDM is a block precoding scheme. One OFDM symbol corresponds to
$N$ frequency-domain information-bearing symbols. By inverse FFT (IFFT), an OFDM symbol is converted into
a block of $N$ time-domain samples. This block is augmented by the cyclic prefix (CP), i.e., by repeating the $L \leq N$ last 
samples at the beginning of the block. 
The OFDM symbol length $N$ and the CP length $L$ are design parameters. 
With CP length $L$, any frequency selective channel with 
impulse response of length $\ell \leq L + 1$ samples is turned into a cyclic convolution channel, 
such that, by applying an FFT at the receiver, it is exactly decomposed into a set of $N$ parallel frequency-flat discrete-time 
channels in the frequency domain. 
Typical CP length is between 16 to 64 time-domain samples. For example, for a 20 MHz signal, as in IEEE 802.11g, 
the time-domain sampling interval is 50 ns, so that a typical CP length ranges between 0.7 and 3.2 $\mu$s.

In a multiuser environment OFDM has also a significant
side advantage: as long as the different users' signals align in time
with an offset not larger than $L - \ell$, where $L$ denotes the CP and $\ell$ is the maximum length of any channel impulse 
response in the system, their symbols after OFDM demodulation remain perfectly aligned in time and frequency.  
In other words, the timing misalignment problem between user signals, which in
single-carrier systems creates significant complications for joint
processing of overlapping signals (e.g., multiuser detection
\cite{verdu_mud}, successive interference cancellation
\cite{wc_book_tse}, Zig-Zag decoding \cite{golakotta08zigzag}),
completely disappears in the case of OFDM, provided that all users
achieve timing alignment within the CP.
 
In a point-to-point MIMO link with $N_r$ receive and $N_t$ transmit antennas, 
the time-domain channel is represented by an $N_r \times N_t$ matrix of 
channel impulse responses. Thanks to OFDM, the channel in the frequency domain
is described by a set of channel matrices of dimension $N_r \times N_t$, one for each of the $N$ 
OFDM subcarriers.  An intuitive explanation of the MIMO multiplexing gain can be given as follows: in the high-SNR regime, the receiver observes 
$N_r$ (noisy) equations with $N_t$ unknown coded modulation symbols on each time-frequency dimension,
each of which carries $\sim \log(\SNR) + O(1)$ bits, where $O(1)$ indicates constants that depend on the channel 
matrix coefficients but are independent of SNR. For sufficiently rich scattering, the rank of the channel matrix is equal to 
$\min\{N_r,N_t\}$ with probability 1. Therefore, using appropriate coding in order to eliminate the effect of the noise, 
up to $\min\{N_r,N_t\}$ symbols per channel time-frequency dimension 
can be recovered with arbitrarily high probability, thus yielding the high-SNR capacity scaling $C(\SNR) = \min\{N_r,N_t\} \log \SNR  + O(1)$ bits/s/Hz. 

{\bf Zero-Forcing Beamforming.} 
In contrast to point-to-point MIMO, in a MU-MIMO system with one $M$-antennas sender and 
$K$ single antenna receivers,~\footnote{We assume single-antenna receivers for simplicity of exposition. 
The extension to $1 \leq N_r \leq M$ antenna receivers is immediate.} it is not generally possible to jointly decode all the receivers observations, 
since the receivers are spatially separated are not generally able to communicate with each other. 
In this case, joint precoding from the transmit antennas must be arranged in order to invert, in some sense, 
the channel matrix and control the multiuser interference.  
One of the techniques to
achieve this is linear Zero-Forcing Beamforming (ZFBF).

In ZFBF, the transmitter multiplies the outgoing symbols
by beamforming vectors such that the receivers see only their intended signals.
For instance, let the received signal on a given OFDM subcarrier at user $k$ be given by
\begin{equation}
y_k = h_{k,1}x_1+h_{k,2}x_2+\dots+h_{k,N_t}x_{N_t}+z_k
\end{equation}
where $h_{k,j}$ is the channel coefficient from transmit antenna $j$ to
user $k$ and $z_k$ is additive white Gaussian noise.
Then, the vector of all received signals can be written in matrix form as
\begin{equation}
\yv = \Hm^\herm \xv+\zv
\label{eqn:channelmodel}
\end{equation}
where $\Hm$ has dimension $M \times K$. 
Assuming $K \leq M$, we wish to find a matrix $\Vm$ such that $\Hm^\herm\Vm$ is zero for all elements
except the main diagonal, that is $\Hm^\herm\Vm=\Lambdam^{1/2} = \diag(\sqrt{\lambda_1},\dots,\sqrt{\lambda_{K}})$. 
Letting $\xv = \Vm \uv$, where $\uv$ is the vector of coded-modulation symbols to be transmitted to the clients, we have
\begin{equation}
\yv = \Hm^\herm \Vm \uv + \zv= \Lambdam^{1/2} \uv + \zv,
\end{equation}
so that each receiver $k$ sees the interference-free Gaussian 
channel $y_k = \sqrt{\lambda_k} u_k + z_k$.

When $\Hm$ has rank $K$ (which is true with probability 1 for sufficiently rich propagation scattering environments
typical of WLANs and for $K \leq M$) a column-normalized version of the 
Moore-Penrose pseudo-inverse generally yields the ZFBF matrix. This takes on the form 
\[ \Vm = \Hm (\Hm^\herm \Hm)^{-1} \Lambdam^{1/2},\] 
where $\Lambdam$ is chosen in order to ensures that the 
norm of each column of $\Vm$ is equal to 1, thus setting the total transmit power
equal to $\trace ( {\rm Cov}(\Vm \uv)) = \EE[\|\uv\|^2]$, i.e., 
equal to the power of the data vector $\uv$.  

As far as the achievable rate is concerned, since ZFBF converts the MU-MIMO channel into a set of 
independent Gaussian channels for each user, subject to the sum-power constraint $\EE[\|\uv\|^2] \leq \SNR$, we have immediately that
the maximum sum rate of ZFBF is given by 
\begin{equation} \label{achievable-zfbf}
R_{\rm sum}^{\rm zfbf}(\SNR) = \sum_{k=1}^K \log \left ( 1 + \lambda_k q_k \right ),
\end{equation}
where $q_k$ denotes the power of the $k$-th data symbol in $\uv$. The above expression can be
maximized over the power allocation $\{q_k\}$, subject to the constraint $\sum_{k=1}^K q_k \leq \SNR$, resulting
in the classical water filling power allocation of parallel Gaussian channels \cite{coverthomas}.

{\bf Tomlinson-Harashima Precoding.} 
In Tomlinson-Harashima Precoding (THP), the mapping
from the data symbol vector $\uv$ to the transmitted symbol vector $\xv$ is {\em non-linear}.
Consider again the channel model (\ref{eqn:channelmodel}).
THP imposes a given precoding ordering, 
and it pre-cancels sequentially the interference of already precoded signals. 
Without loss of generality, consider the natural precoding ordering to be from $1$ to $K$.
Let $\Hm = \Qm \Rm$ be the QR factorization of $\Hm$,  such that $\Rm$ is $K \times K$ upper triangular 
with real non-negative diagonal coefficients, and $\Qm$ is $M \times K$ tall unitary, 
such that $\Qm^\herm \Qm = \Id$.  THP precoding is formed by the concatenation of a linear mapping, 
defined by the unitary matrix $\Qm$,  with a non-linear mapping that does the interference pre-cancellation. 
Let $\hat{\uv} = {\rm THP}(\uv)$ denote the  non-linear mapping of the data vector $\uv$ into an intermediate vector 
$\hat{\uv}$, that will be defined later.  The linear mapping component of THP is then given by 
\begin{equation} 
\xv = \Qm \hat{\uv}, 
\end{equation}
where ${\rm Cov}(\hat{\uv}) = \Sigmam = \diag(q_1,\ldots, q_K)$ 
and, as before, $q_k$ denotes the power allocated to the $k$-th data symbol.  
It follows that the channel reduces to
\begin{eqnarray} \label{marameo1}
\yv  & = &  \Hm^\herm \xv + \zv \nonumber \\
& = & \Rm^\herm \Qm^\herm \Qm \hat{\uv} + \zv \nonumber\\
& = & \Lm \hat{\uv} + \zv, 
\end{eqnarray}
where $\Lm = \Rm^\herm$ is lower triangular. 
The signal seen at client $k$ receiver is given by 
\begin{equation} \label{channel-without-precoding} 
y_k = [\Lm]_{k,k} \hat{u}_k + \underbrace{\sum_{j < k} [\Lm]_{k,j} \hat{u}_j}_{\mbox{interference}} + z_k. 
\end{equation}
Next, we look at the non-linear mapping $\uv \mapsto \hat{\uv}$. The goal is to pre-cancel the term 
indicated by ``interference'' in (\ref{channel-without-precoding}). 
Notice that this term depends only on symbols $\hat{u}_j$ with $j < k$. Therefore, the elements $\hat{u}_1, \ldots, \hat{u}_K$ can be 
calculated sequentially.  A simple pre-subtraction of the interference term at each step would increase the effective transmit power and
and would result in a suboptimal version of the linear ZFBF treated before.

The key idea of THP is to introduce a modulo operation that limits the transmit power of each precoded stream
$\hat{u}_k$. This is defined as follows. Assume that the data symbols $u_k$ are points from a QAM constellation 
uniformly spaced in the squared region of the complex plane bounded by the interval $[-\tau/2, \tau/2]$ on both the real axis 
and the imaginary axis. 
Then, for a complex number $s$, let $s$ modulo $\tau$ be given by 
$[s]_{{\rm mod} \; \tau} = s - Q_\tau(s)$, where $Q_\tau(s)$ is the point $(n + jm)\tau$ with integers $n,m$ closest to $s$. 
In short, $Q_\tau(s)$ is the quantization of $s$ with respect to a square grid with minimum distance $\tau$ on the complex plane, 
and $[s]_{{\rm mod} \; \tau}$ is the quantization error. We let
\begin{equation} \label{minchiata}
\hat{u}_k = \sqrt{q_k} \left [ u_k - \frac{\sum_{j < k} [\Lm]_{k,j} \hat{u}_j}{[\Lm]_{k,k} \sqrt{q_k}} \right ]_{{\rm mod} \; \tau}.
\end{equation}
In this way, the symbol $\hat{u}_k$ is necessarily bounded into the squared region of side $\tau \sqrt{q_k}$, 
and its variance (assuming a uniform distribution over the squared region, which is approximately true when we use a QAM constellation inscribed in the square) is given by $\EE[|\hat{u_k}|^2] = \tau^2/6 q_k$. Letting $\tau = \sqrt{6}$ we have that the precoded symbols have
the desired power $q_k$. 

Let's focus now on receiver $k$ and see how the modulo precoding can be undone. 
The receiver scales the received symbol $y_k$ by $[\Lm]_{k,k} \sqrt{q_k}$ and applies again the same the modulo $\tau$ non-linear mapping. 
Simple algebra then shows that 
\begin{eqnarray} \label{thp-receiver}
\widehat{y}_k & = & \left [ u_k + \frac{z_k}{[\Lm]_{k,k} \sqrt{q_k} } \right ]_{{\rm mod} \; \tau}.
\end{eqnarray}
It follows that the interference term is perfectly removed, but we have introduced a distortion in the noise term. 
Namely, while $u_k$ is unchanged by the modulo operation, since by construction it is a point inside the square, 
the noise term $\frac{z_k}{[\Lm]_{k,k} \sqrt{q_k} }$ is ``folded'' by the modulo operation, i.e., the tails of the Gaussian noise distribution
are folded on the squared region. Noise folding is a well-known effect of THP \cite{forney-eyuboglu}. 

As far as the achievable rate is concerned, it is possible to show (see \cite{tosato-boccardi-caire,erez-zamir-shamai}) that this is given by 
\begin{equation} \label{achievable-lowerbound-thp-i}
R_{\rm sum}^{\rm thp}(\SNR) = \sum_{k=1}^K \left [ \log(1 + |[\Lm]_{k,k}|^2 q_k) - \log (\pi e/6) \right ]_+,
\end{equation}
where $[\cdot]_+$ indicates the positive part. Again, this sum rate can be optimized with respect to the power allocation
$\{q_k\}$, subject to the sum power constraint $\sum_{k=1}^K q_k \leq \SNR$. 
The rate penalty term $\log (\pi e/6)$ is the shaping loss, due to the fact that THP produces a signal which is uniformly 
distributed in the square region (therefore, a codeword of $n$ signal components is uniformly distributed in an $n$-dimensional 
complex hypercube).~\footnote{It should be noticed that the same shaping loss at high SNR is incurred by any other 
scheme, including plain CSMA, when practical QAM constellations are used instead of 
the theoretically optimal Gaussian coding ensemble.}

\section{Synchronization in distributed MIMO systems}
\label{sec:idea}

In a distributed MU-MIMO setting, 
timing and carrier phase synchronization across the jointly precoded APs 
are needed in order for ZFBF and THP precoding to work. 
As discussed above, timing synchronization requires only that all nodes align their slots within the length of the 
OFDM CP. This is relatively easy to achieve, and it has already implemented in software radio testbeds as in 
\cite{rahul10sourcesync, tan10fine}. 
Carrier phase synchronization, however, is much more challenging. 
While a centralized MU-MIMO transmitter has a common clock source for all 
its RF chains \cite{argos}, 
in a distributed setting each AP has an individual clock. 
The relative time-varying instantaneous phase offset between the different transmitters may cause 
a phase rotation of the transmitter signals across a downlink slot such that, even though at the beginning of the slot
we have ideal precoding (e.g., ZFBF or THP), the interference nulling effect is completely destroyed towards the end of the slot. 

It is important to remark here that, while synchronizing a receiver with a transmitter for the purpose of coherent detection is a well-known problem for which robust and efficient solutions
exist and are currently implemented in any coherent digital receiver \cite{proakis}, 
here we are faced with a different and significantly harder problem, which consists of synchronizing the instantaneous 
carrier phase of different {\em transmitters}. 
This requires that APs must track an RF carrier reference and compensate for the relative (time-varying) phase rotation
while they are transmitting the downlink slot. Simultaneous transmission of the data signal and reception of the carrier reference signal
cannot be implemented by standard off-the-shelf terminals. Instead, we have devised a system architecture to accomplish this goal. 

\begin{figure*}[!b]
% ensure that we have normalsize text
\normalsize
% Store the current equation number.
\setcounter{MYtempeqncnt}{\value{equation}}
% Set the equation number to one less than the one
% desired for the first equation here.
% The value here will have to changed if equations
% are added or removed prior to the place these
% equations are referenced in the main text.
\setcounter{equation}{12}
\vspace*{3pt}
\hrulefill
{\small
\begin{equation} \label{eq:effective} 
\widetilde{\Hm} (n; t) = 
\underbrace{\left [ \begin{array}{cc} 
e^{j (\frac{2\pi}{NT_s} \tau_1 n + \phi_1(t))} & 0 \\ 0 & e^{j (\frac{2\pi}{NT_s}\tau_2 n + \phi_2(t))} \end{array} \right ]}_{\Phim(n;t)} 
\left [ \begin{array}{cc} 
H_{11}(n) & H_{12}(n) \\ 
H_{21}(n) & H_{22}(n) \end{array} \right ] 
\underbrace{\left [ \begin{array}{cc} 
e^{j (\frac{2\pi}{NT_s} \delta_1 n + \theta_1(t))} & 0 \\ 0 & e^{j (\frac{2\pi}{NT_s} \delta_2 n + \theta_2(t))} \end{array} \right ]}_{\Thetam(n;t)} 
\end{equation}
} 
% Restore the current equation number.
\setcounter{equation}{\value{MYtempeqncnt}}
% IEEE uses as a separator
%\hrulefill
% The spacer can be tweaked to stop underfull vboxes.
%\vspace*{4pt}
\end{figure*}

{\bf Why is distributed MU-MIMO challenging?} 
For simplicity of exposition, consider a distributed MU-MIMO system with
two clients and two access points, each one with a single antenna and using ZFBF
(the following considerations apply immediately to more general scenarios).
For nomadic users, as in typical  WLAN setting, 
the physical propagation channel changes quite slowly with time, 
so that we may assume that the channel impulse response is locally time-invariant.
In order to use ZFBF, the channel matrix coefficients at each OFDM 
subcarrier must be estimated and known to the transmitter central server.
Various methods for learning the downlink channel matrix at the transmitter side 
have been proposed, including closed-loop feedback schemes (see \cite{caire2010multiuser} and the references therein) 
or open-loop schemes that exploit the uplink/downlink channel reciprocity of TDD systems \cite{jose2008scheduling}. 
For simplicity of exposition, we will assume here that the channel estimates correspond perfectly 
to the actual channel.

The central server computes the precoding matrix as seen in in Section \ref{sec:primer}, for each subcarrier 
$n = 1, \ldots, N$. Let
\begin{equation} \label{nominal} 
\Hm(n) = \left [ \begin{array}{cc} 
H_{11}(n) & H_{12}(n) \\ 
H_{21}(n) & H_{22}(n) \end{array} \right ] 
\end{equation}
denote the $2 \times 2$ downlink channel matrix between 
the two clients and the two access point antennas on subcarrier $n$, and let 
$\Vm(n)$ denote the corresponding precoding matrix such that 
$\Hm^\herm (n) \Vm(n) = \Lambdam^{1/2}(n)$ is diagonal. 
If the timing and carrier phase reference remain unchanged from
the slot over which the channel is estimated and the slot over which the precoded signal is transmitted, we obtain perfect zero-forcing of the 
multiuser interference.

Suppose now that the timing reference and carrier phase reference
between the estimation and transmission slots of the two APs
is not ideal.  With perfect timing, the downlink channel from AP $i$ to
client $j$ would have impulse response $h_{ij}(\tau)$.  Instead, due to lack of synchronization, the impulse response is 
$h_{ij}(\tau - \tau_i - \delta_j) e^{j(\phi_i(t) + \theta_j(t))}$ where $\tau_i, \delta_j$ denote the timing misalignment 
of AP $i$ and client $j$, respectively, and  $\phi_i(t), \theta_j(t)$ 
denote the instantaneous phase differences (with respect to the nominal RF carrier reference)
of AP $i$ and client $j$, respectively.  
For simplicity, we assume here that the sampling clock at all nodes is precise enough such that we may assume
that the sampling frequency is the same and does not change significantly in time over the duration of a slot. 
Hence, $\tau_i$ and $\delta_j$ can be considered as unknown constants. 
Furthermore, we assume that they are multiples of the sampling interval $T_s$ (i.e., the duration of the time-domain samples) 
otherwise the derivation is more complicated, involving the folded spectrum of the channel frequency 
response, but the end result is equivalent to what derived here. 
Instead, we model the instantaneous phases of the RF carrier oscillators as
\begin{eqnarray} \label{ziofanale}
\phi_i(t) & = & \phi_i(0) + 2\pi \Delta_i t + w_i(t) \nonumber \\
\theta_j(t) & = & \theta_j(0) + 2\pi \Delta_j t + w_j(t)
\end{eqnarray}
where $\phi_i(0), \theta_j(0)$ are unknown constants, $\Delta_i = (f_{c,i} - f_c) (N+L)T_s$ is the normalized frequency offset of
node $i$ with respect to the nominal carrier frequency $f_c$, and $w_i(t)$ is a zero-mean stationary phase noise process, 
whose statistics depends on the hardware implementation. In the above expression, the time index $t$ ticks at the OFDM symbol rate, 
i.e., at intervals of duration $(N+L)T_s$. 

From the well-known rules of linearity and time-shift of the discrete
Fourier transform, we arrive at the expression for the effective
channel matrix in (\ref{eq:effective}).
The diagonal matrix of phasors 
$\Phim(n;t)$ and $\Thetam(n;t)$ depend, in general, on both the subcarrier and OFDM symbol indices $n$ and $t$. 
The multiplication of the nominal channel matrix $\Hm(n)$ from the right (receiver side, according to the channel model
(\ref{eqn:channelmodel})) poses no problems, since these phase shifts can be recovered individually by 
each client as in standard coherent communication \cite{proakis}. In contrast, the diagonal matrix $\Phim(n)$ 
multiplying from the left (the transmitter side) poses a significant problem: 
since the server computes the MIMO precoding matrix $\Vm(n)$ based on its estimate $\Hm(n)$, 
it follows that when applied to the effective channel $\widetilde{\Hm}(n)$ in (\ref{eq:effective}) 
the matrix multiplication $\widetilde{\Hm}^\herm(n) \Vm(n)$ is in general far from diagonal. 
To stress the importance of this aspect, we would like to make
clear that the resulting signal mixing takes place over the actual
transmission channel, making it impossible for the receivers to eliminate it.

{\bf Why Synchronization Is Possible.} 
Any discussion on phase synchronization of distributed wireless
transmitters must necessarily start with the mechanisms through which
phase errors occur. Digital wireless transmission systems are
constructed using a number of clock sources, among
which the two most important ones are the sampling clock and the
carrier clock. In a typical system, signals are created in a digital
form in baseband at a sampling rate on the order of tens of MHz, then
passed through a digital-to-analog converter (DAC). Through the use of
interpolators and filters, the DAC creates a smooth analog waveform
signal which is then multiplied by a sinusoidal carrier produced by the
carrier clock. The result is a passband signal at a frequence of a few GHz which is then sent over
the antenna.

Wireless receivers, in turn, use a chain of signal multiplications and
filters to create a baseband version of the passband signal
received over the antenna. Some designs, such as the common
superheterodyne receivers, use multiple high frequency clocks and
convert a signal first to an intermediate frequency before bringing it
back to baseband. Other designs simply use a carrier clock operating
at the same nominal frequency as the carrier clock of the transmitter
and perform the passage from passband to baseband in a single step. We
will be focusing on such designs in the ensuing discussion. After
baseband conversion, the signal is sampled and the resulting digital
waveform is decoded.

There are four clocks in the signal path: the transmitter's sampling
clock and RF carrier clock and the receiver's RF carrier clock and sampling
clock. All four clocks manifest phase ``drift'' (i.e., a linear time-varying term) 
and ``jitter'' (i.e., a random fluctuation term). 
We have assumed that the sampling clocks have no significant drift and jitter, and
the only effect of timing misalignment (within the length of a CP) is captured by the 
constants $\tau_i$ and $\delta_j$ in (\ref{eq:effective}). In contrast, 
the carrier clocks are affected both by drift and jitter (see (\ref{ziofanale})). 
Furthermore, the phase noise term $w_i(t)$ may have some slow dynamics that can be linearized locally, 
over the duration of a slot,  and add up to the linear phase term, such that the slope of the phase drift 
is constant over a single slot, but it is not constant over longer time intervals, in general. 

\begin{figure}
 \centering
 \includegraphics[width=.4\textwidth]{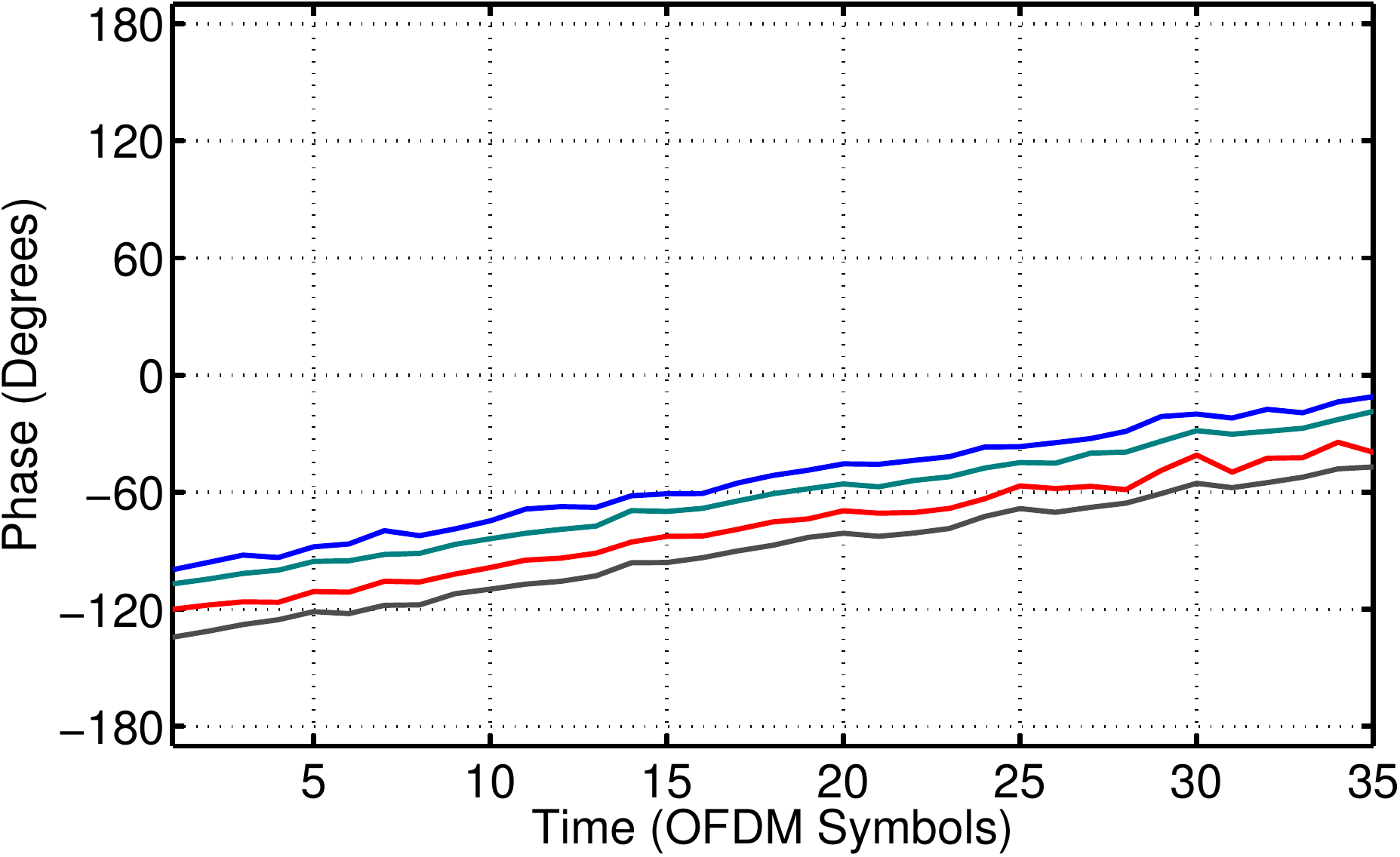}
 \caption{Pilot phases}
 \label{pilots}
\end{figure}

We have verified experimentally the validity of our model, by letting a transmitter 
send several tone signals, i.e.,  simple unmodulated sine waves, 
corresponding to different subcarriers of the OFDM modulation, and using a receiver to sample, demodulate
and extract the instantaneous phase trajectory of the received tones. 
In the absence of phase offset these signals would exhibit a constant phase when measured over a sequence of 
several OFDM symbols. Instead,  the measured instantaneous phase is time-varying and closely approximate
parallel straight lines, as shown in  Figure \ref{pilots}. The common slope of these straight lines
is given by the carrier frequency offset $\Delta_i$ between transmitter and receiver. 
The spacing between the lines is given by constant phase terms $\frac{2\pi}{NT_s}\tau_i n$ for different subcarrier index $n$, 
and depends on the time misalignment $\tau_i$ between the AP and 
the nominal slot initial time. The small fluctuations around the 
linear behavior of the instantaneous phase is due to the phase noise, which 
is quite small for the WARP hardware used in our system, as it can be observed qualitatively from 
plots as in Figure \ref{pilots}.

It follows that by estimating the spacing between the phase trajectories (intercepts with the horizontal axis) and 
their common slope, we can track and predict across the slot the phase de-rotation coefficients to be applied 
at each AP in order to ``undo'' the effect of the matrix $\Phim(n;t)$. Notice that the de-rotation factor must 
be predicted a few OFDM symbols ahead, in order to include the delay of the hardware implementation between 
when an OFDM symbol is produced by the baseband processor (FPGA) to when it is actually transmitted. 

%% file: system.tex
\section{AirSync}
\label{sec:hardware}

The fact that the common phase drift of all subcarriers can be
predicted by observing only a few of them prompts the following
approach to achieving phase synchronization between access points: a
main access point (master) is chosen to transmit a reference signal
consisting of several pilot tones \footnote {The use of multiple pilot
  tones ensures frequency diversity and spreads the pilot signal power
  over multiple frequency bins.}  placed outside the data transmission
band, in a reserved portion of the available bandwidth. An initial
channel probing header, transmitted by the master access point, is
used by the other transmitters in order to get an initial phase
estimate for each carrier. After this initial estimate is obtained,
the phase estimates will be updated using the phase drift measured by
tracking the pilot signals. The estimate is used to calculate the
difference between the carrier phase of each secondary transmitter and
the phase of the master transmitter. This difference depends on the
timing offset between the starting points of their frames and the
frequency offset between the carrier frequency of the master AP
(denoted by $f_{c,1}$) and the carrier frequency of each secondary AP
(denoted by $f_{c,i}$, for $i > 1$). After obtaining the channel
estimate, the secondary transmitters are able to undo effect of the
instantaneous phase difference by derotating the transmitted
frequency-domain symbols by the phase difference term along the whole
transmission slot, thus eliminating the presence of the time-varying
diagonal matrix $\Thetam(n;t)$ in front of the estimated channel
matrix and therefore achieving the desired MU-MIMO precoding along the
whole transmission slot.

More specifically, at time $ t = 0 $, the $n$-th subcarrier signal
generated by the master AP has the phase $ \frac{2\pi}{NT_s} \tau_1 n
+ \phi_1(0) $, while the carrier generated by AP $ i $ has the phase $
\frac{2\pi}{NT_s} \tau_i n + \phi_i(0) $. The phase of the
instantaneous phase difference obtained from the master pilot tones
is, ignoring the phase noise terms, $ \frac{2\pi}{NT_s} (\tau_1 -
\tau_i) n + \phi_1(0) - \phi_i(0) + \angle H_i(n) $, where $ \angle
H_i(n) $ is the phase of the channel coefficient between the master AP
and AP $ i $. If this phase estimate is added to the phase of the
generated $n$-th subcarrier at AP $ i $, the resulting phase becomes $
\frac{2\pi}{NT_s} \tau_1 n + \phi_1(0) + \angle H_i(n) $, that is the
phase of AP $ i $ is the phase of the master AP plus an offset $
\angle H_i(n) $. To keep this offset constant over the duration of a
transmission slot, the estimate must be adjusted by adding, for all
$t$ ranging over the transmission slot, the linear relative phase
drift term $2\pi (\Delta_1 - \Delta_i) t$.  In this way, after the
phase compensation, all APs transmit at the {\em actual} frequency
$f_{c,1}$ of the master AP.

The drift $2\pi (\Delta_1 - \Delta_i) t$ is estimated based on the
out-of-band pilots using a sliding window smoothing filter over four
samples to compute an updated value of the ``slope'' $\Delta_i -
\Delta_1$. The secondary AP predicts, based on the current estimate,
the instantaneous phase with a few OFDM symbols of look-ahead. The
need for look-ahead prediction arises from the fact that the AP must
align its phase to the phase of the reference at the moment of the
actual transmission, not at the moment that the estimate has been
recorded.  Thus the look-ahead time of $ d $ OFDM symbols corresponds
to the synchronization circuit delay. The prediction is obtained by
simple linear extrapolation, by letting the correction term at time $t
+ d$ be given by $2\pi (\Delta_1 - \Delta_i) (t + d)$, where $\Delta_1
- \Delta_i$ is the estimated slope at time $t$. The constant offset
$\angle H_i(n)$ becomes a part of the downlink channel estimates and
poses no further problems with regard to synchronization both when
using downlink and uplink channel estimation schemes.

In our current implementation, for simplicity, we obtain an individual
phase estimate of the form $ \frac{2\pi}{NT_s} (\tau_1 - \tau_i) n +
\phi_1(0) - \phi_i(0) + \angle H_i(n) $ for every subcarrier and use
it independently of the estimates for other subcarriers in correcting
the subcarrier phase. The form of the phase estimate suggests that it
is possible to obtain a better estimate by breaking the estimation
process into two distinct parts: obtaining an initial, high quality
estimate of the constant $ \angle H_i(n) $ during a system calibration
step and then estimating just the two factors $ \tau_1 - \tau_i $ and
$ \phi_1(0) - \phi_i (0) $ in subsequent packet transmissions. The
constant estimate in this case is needed since undoing the angle $
\angle H_i(n) $ amounts to equalizing the channel between the master
AP and the $i$-th AP. After equalizing the channel, the resulting
phases can be unwrapped along the carrier index $ n $. It results
that, after compensating for the angle $ \angle H_i $, the phase of
the estimate is $ \frac{2\pi}{NT_s} (\tau_1 - \tau_i) n + \phi_1(0) -
\phi_i(0) $, linear in the carrier index plus a constant term.  A
linear MMSE fitting can be applied in order to find the two factors
mentioned, which are in fact the slope of the line (the carrier phase
with regard to the subcarrier index) and its intercept.

{\bf Software Radio Implementation.} We have implemented AirSync as a
digital circuit in the FPGA of the WARP radio platform
\cite{warp}. The WARP radio is a modular software radio platform
composed of a central motherboard hosting an FPGA and several
daughterboards containing radio frequency (RF) front-ends. The entire
timing of the platform is derived from only two reference oscillators,
hosted on a separate clock board: a 20 MHz oscillator serving as a
source for all sampling signals and a 40 MHz oscillator which feeds
the carrier clock inputs of the transceivers present on the RF
front-ends. The shared clocks assure that all signals sent and
received using the different front-ends are phase synchronous. Phase
synchronicity for all sent signals or for all received signals is a
common characteristic of MIMO systems. However, the fact that the
design of the WARP ensures phase synchronicity among the sent and
received signals, as opposed to using separate oscillators for
modulation and demodulation, greatly simplifies the synchronization
task. The system's data bandwidth is 5 MHz. We place the
synchronization tones outside the data bandwidth, at about 7.5 MHz
above and below the carrier frequency. 

The slave APs have to track the out of band pilots (i.e., receive
these signals) and transmit the data signal \emph{at the same time},
in an FDD manner. We have dedicated one antenna of each secondary AP
to receiving and tracking the reference signal, while the other
antennas are used for transmitting phase-synchronous signals. The
system design must mitigate self-interference between the transmit and
receive paths. 

In FDD transmission schemes in which the front-ends sample the entire
system bandwidth, the dynamic ranges of the ADC and DAC circuitry
plays an important limiting role. As opposed to a complete full-duplex
system, in which self-interference cancellation is the main challenge
to be solved, in bandwidth sharing systems the main challenge is
accommodating both the incoming signal, i.e. the signal from the
master AP, and the secondary AP's data signal within the limited
dynamic range of the secondary's receiver front-end. A second
challenge is shaping this data signal in order to prevent any
significant power leakage outside the data band, mitigating the need
for large guard bands between the data and the pilots.

The dynamic range needed can be computed as follows: assume that the
secondary AP's signal and the master AP's pilots are broadcasted at
the same power level. If the secondary's receiver antenna is $ \alpha
$ times closer to the secondary's transmitter antenna than to the one
of the master's, assuming a free space propagation model in which the
power decays as $\frac{1}{\alpha^2}$, it results that the data signal
is received at $ 10*\log_{10}(\alpha^2) $ dB above the pilot
signals. For $ \alpha $ in the 32 to 128 range, this amounts to 30 dB
to 42 dB. For comparison, the WARP's 14-bit ADCs offer a dynamic range
of 84 dB. \footnote{This requirement could be further relaxed through
  the use of an analog rejection filter over the data band, before
  sampling, during the tracking period, thus decreasing the needed
  dynamic range through receive-side filtering.}

For the second problem, the design of WiFi-NC
\cite{chintalapudi12wifi}, offers a clear indication of what can be
achieved in a software radio using the same components as the WARP.
To limit the size of required guard bands in a bandwidth-sharing
system, in which different APs divide the data band into slices and
can transmit in duplex over separate slices, the authors construct an
OFDM transmitter with a sharp spectral footprint. By employing digital
filters in the FPGA, they achieve a 60 dB power decay with guard bands
that total 4\% of the data bandwidth, as proved by spectrum analyzer
plots. Their filter response time is well within the cyclic
prefix. This approach allows for decreasing the over-the-channel power
leakage into the pilot band through sender-side filtering. In our
system, we achieve a similar effect by using the baseband sender filter
present in the transmit signal path of the WARP's transceiver. In
general, self-interference can been avoided using a number of other
techniques such as antenna placement \cite{katti}, digital
compensation \cite{ashutosh}, or simply relying on the OFDMA-like
property of a symbol aligned system \cite{tan10fine} and preventing
the secondary APs from using the pilot subcarriers.

\begin{figure}
 \centering
 \includegraphics[width=.4\textwidth]{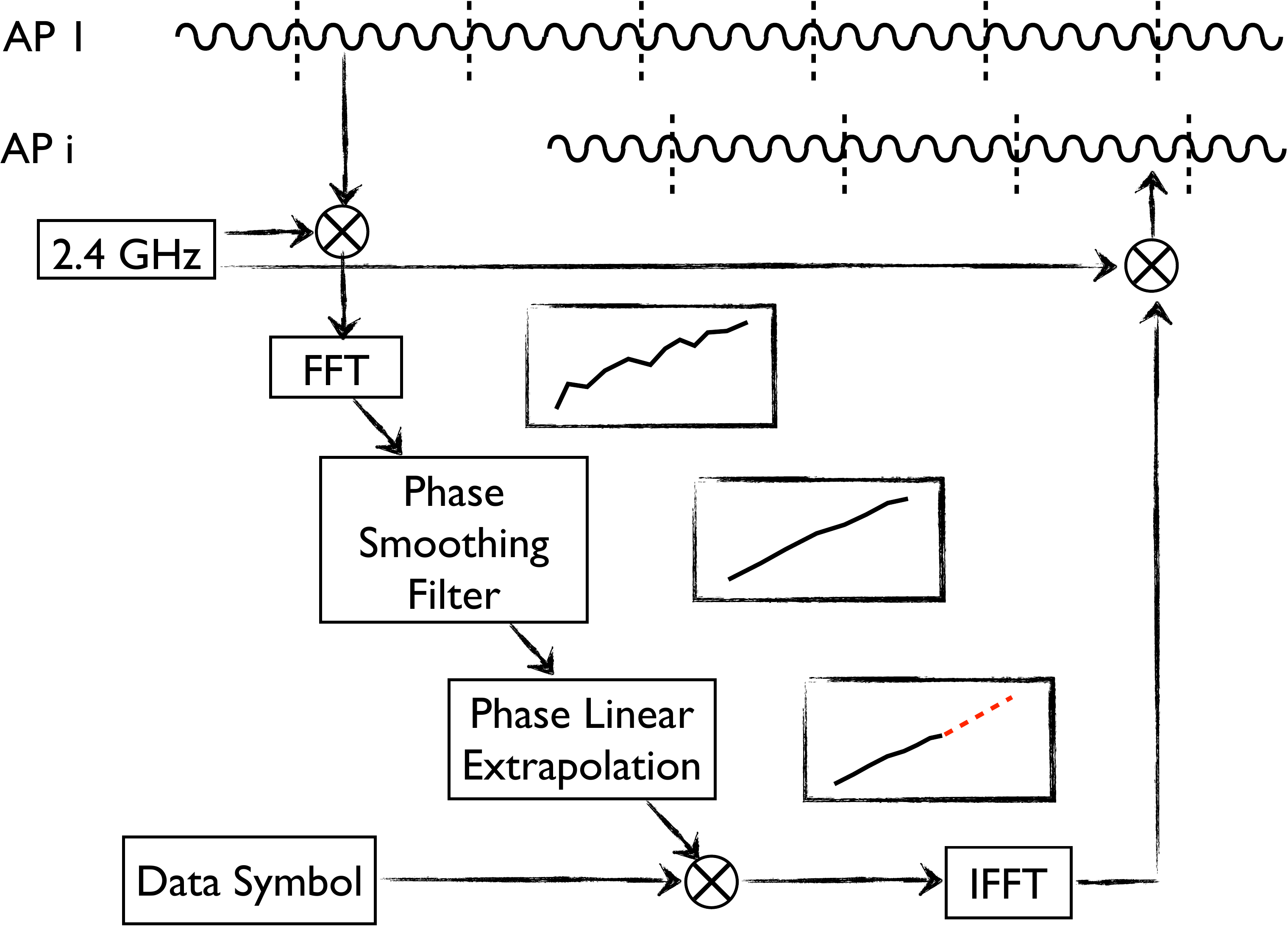}
 \label{fig:airsync}
 \caption{Airsync operation: a secondary AP (bottom) synchronizes its
   phase to the one of a reference signal (top) by adjusting the phase of
   its signal to match the phase of the reference.}
\end{figure}

We have implemented a complete system-on-chip design in the FPGA,
taking advantage of the presence of hard-coded ASIC cores such as a
PowerPC processor, a memory controller capable of supporting transfers
through direct memory access over wide data buses and a gigabit
Ethernet controller. Atop this system-on-chip architecture we have
ported the NetBSD operating system and created drivers for all the
hardware components hosted on the platform, capable of setting all
system and radio board configuration parameters. The operating system
runs locally but mounts a remote root filesystem through NFS. In the
same system-on-chip architecture we integrated a signal processing
component created in Simulink which provides interfaces for fast
direct memory access. This latter component is responsible for all the
waveform processing and for the synthesis of a phase synchronous
signal and interfaces directly with the digital ports of the radio
front-ends. We interfaced the Ethernet controller and the signal
processing component using an operating system kernel extension
responsible for performing zero-copy, direct memory access data
transfers between the two, with the purpose of passing back and forth
waveform data at high rates between a host machine and the WARP
platform. The large data rates needed (160 Mbps for a 5MHz wireless
signal sampled at the 16 bit precision of the WARP DACs for both the
real (I) and imaginary (Q) parts of the corresponding baseband signal)
required optimizing the packet transfers into and out of the WARP.
For example, consider the direct memory access ring associated with
the receive end of the Ethernet controller on the board, which is
shared between packets destined to the signal processing component and
packets destined to the upper layers of the operating system stack. We
do not release and reallocate the memory buffers occupied by packets
destined to the signal processing component. Instead, we use a lazy
garbage collection algorithm in order to reclaim these buffers when
they are consumed in a timely manner or reallocate them at a later
point if they are not consumed before the memory ring runs low on
available memory buffers. The rationale for this particular
optimization is that the overhead of managing the virtual-memory based
reallocation of memory buffers of tens of thousands of packets every
second would bring the processor of the software radio platform to a
halt.

All transmitting WARP radios are connected to a central processing
server through individual Ethernet connections operating at gigabit
speeds. Most of the signal synthesis for the packet transmission is
done offline, using Matlab code. We produce precoded packets in the
form of frequency domain soft symbols. However, the synchronization
step and the subsequent signal generation is left to the FPGA. The
server, a fast machine with 32 processor cores and 64GB of RAM,
encodes the transmitted packets and streams the resulting waveforms to
the radios.
Figure \ref{fig:airsync} illustrates the process of creating a phase
synchronous signal at the secondary AP.

{\bf Centralized joint encoding.} By transmitting phase synchronous
signals from multiple APs, we have created a virtual single MU-MIMO
transmitter, for which standard MU-MIMO precoding strategies can be
used.  However, the use of distributed APs complicates the design of
the transmitter system.  In order to eliminate multiuser interference,
the data streams to different clients must be jointly precoded, as we
have seen in Section \ref{sec:primer}.  For systems with a very large
number of jointly processed antennas and targeting mobile cellular
communications (e.g., see \cite{argos}), the
centralized computation of the precoding matrix, of the precoded based
band signals, and distribution of these signals to all the antennas
would require a large delay, which is incompatible with the short
channel coherence time due to user mobility.  In contrast, in our
enterprise network or residential network scenario, the channel
coherence time is much longer (typical users are nomadic, and move at
most at walking speed). Therefore, computing the precoding matrix does
not represent a significant problem, and it is in fact better to
perform centralized precoding and distribution of the baseband
precoded signals.  For example, using the conjugate beamforming scheme
of \cite{argos}, it is possible to compute the
precoded signals in a decentralized way, since each AP $i$ needs just
to combine the clients' data streams with the complex conjugates of
its own estimated channel coefficients, i.e., with the elements of the
$i$-th row of the channel matrix. In the notation of Section
\ref{sec:primer}, this corresponds to letting $\xv = c \Hm \uv$, for
some power normalizing constant $c$, such that the precoded channel
becomes $\yv = c \Hm^\herm \Hm \uv + \zv$. Unless $M \gg K$, the
resulting matrix $\Hm^\herm \Hm$ is far from diagonal, and the system
is interference limited, i.e., by increasing the transmit power, the
system sum rate saturates to some constant value (the system
multiplexing gain in this case is 1, corresponding to serving only one
client on each time-frequency dimension, as in standard FDMA/TDMA).  Hence,
while conjugate beamforming is an attractive scheme for very large
$M$, relatively high client mobility and limited power (as in a
cellular system), it turns out that in the WLAN setting with not so
large $M$, low client mobility and large operating SNR (due to
communication range of at most a few tens of meters) this is not a
competitive choice.

As a matter of fact, centralized ZFBF or THP precoding is much better
in our setting. It should also be noticed that by centralized
precoding we need only to send the I and Q components of the
frequency-domain OFDM baseband (precoded) symbols to the APs. This
requires roughly $2 b \times W$ bit/s, for signal bandwidth $W$ Hz and
$b$ quantization bits per real sample. Instead, decentralized
processing requires to send {\em all} client data streams to all
APs. Assume for example that $ K $ clients are receiving at 4 bit/s/Hz
(corresponding to 20 Mbps over a $W = 5$ MHz bandwidth). This requires
$20 \times K$ Mbps to be sent to all APs, while in the case of
centralized processing, with $b = 16$ bits of quantization, we need
only $32 \times 5$ Mbps. Here, for $K > 5$, centralized processing is
convenient also in terms of the backhaul data rate. For sufficiently
large $K$, centralized processing is eventually less demanding than
decentralized processing in terms of the backhaul data rates.

Our central server has an individual gigabit Ethernet connection to
each of the WARP radios serving as APs. We divide the
downlink time into slots and in each slot schedule for transmission a
number of packets destined to various clients, 
according to an algorithm that will be presented in Section \ref{section:medium}. 
For each AP, the server computes the I and Q components of the precoded baseband 
frequency domain waveform to be transmitted in the next downlink slot. 
However, it does not perform any phase correction at this point. 
The only information used in the precoding is the data to be transmitted 
and the channel state information between APs and clients. 
The server transmits their corresponding waveforms to all secondary APs, 
and finishes by feeding the master AP, so that the master AP 
starts transmitting right away and the secondary AP can immediately synchronize and 
follow.

At the moment we obtain CSI using a downlink estimation procedure,
similar to the one presented in 802.11ac. In a future refinement of
our system, we would like to reduce the overhead of obtaining CSI by using
an uplink estimation scheme that takes advantage of channel
reciprocity, thereby reducing considerably  the length of the channel estimation
procedure.

%% file: performance.tex
\section{Performance Evaluation}
\label{section:performance}

\begin{figure}[t]
 \centering
 \includegraphics[width=.45\textwidth]{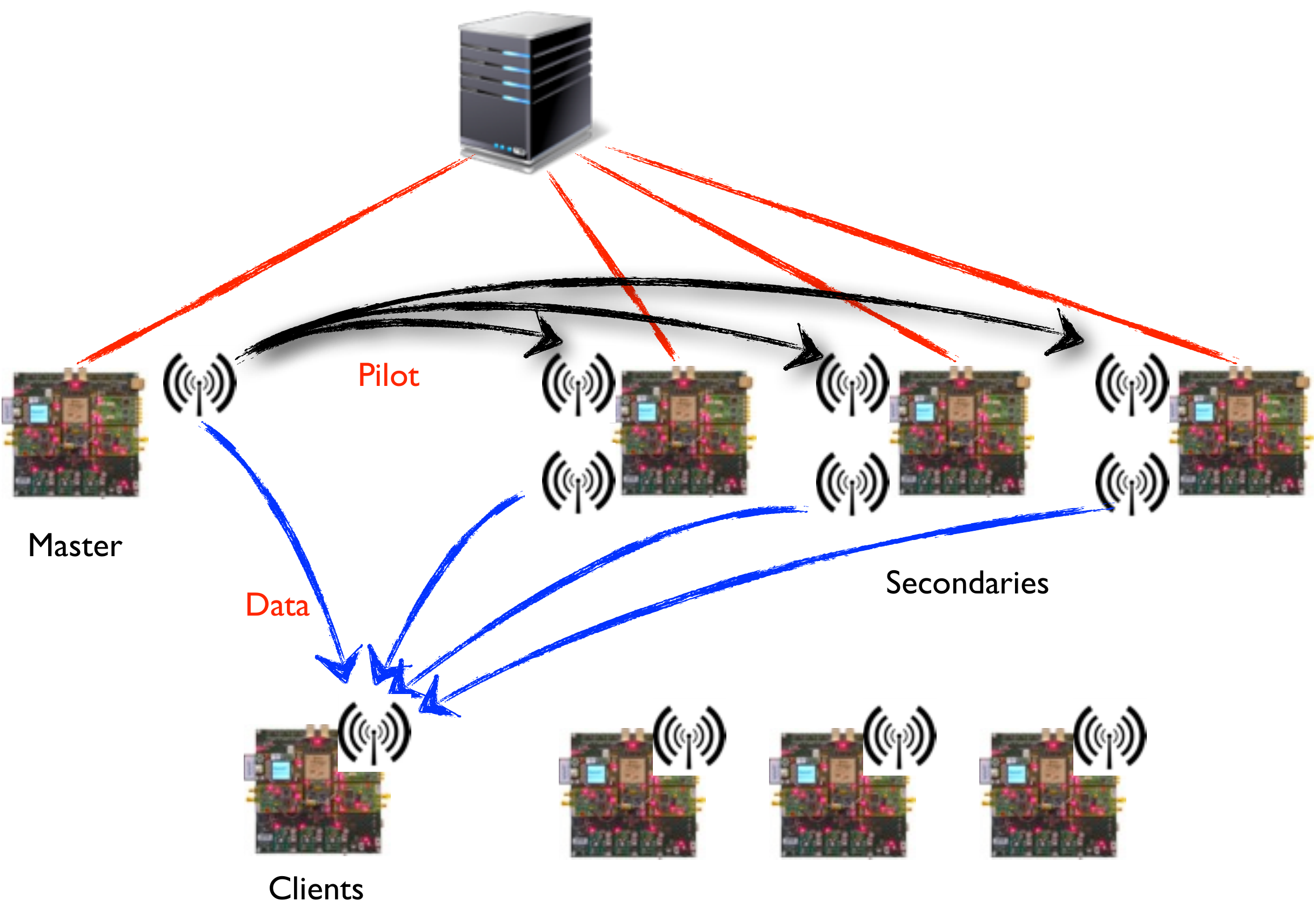}
 \caption{Testbed diagram. \textnormal{The central server is connected to four transmitters, the main transmitter on the left and the three secondary transmitters on the right. Four receivers act as clients.}}
 \label{testbed}
\end{figure}

\begin{figure*}[t]
 \centering
 \includegraphics[width=.98\textwidth]{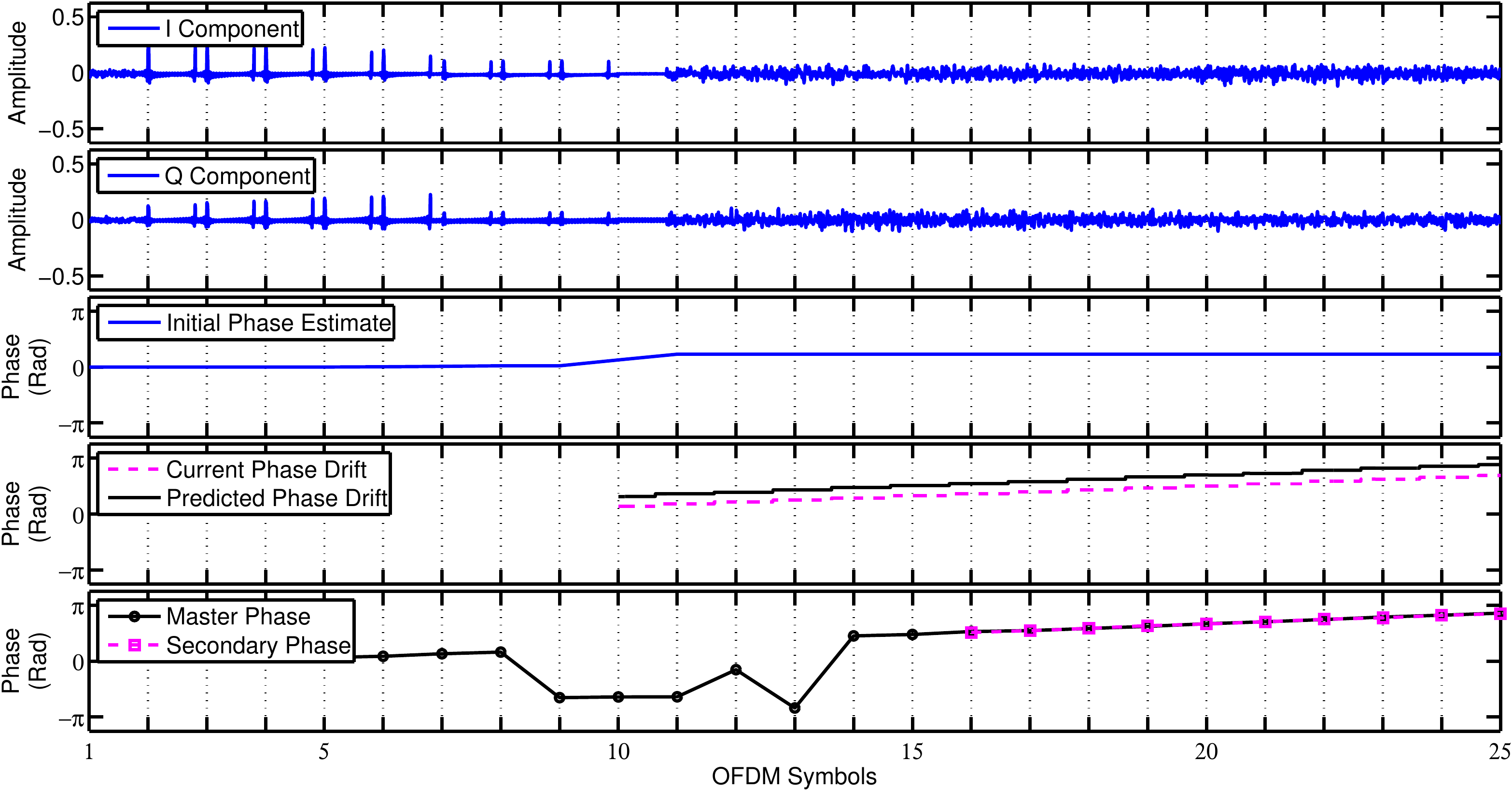}

 \caption{Phase Synchronization Acquisition. \textnormal{The secondary transmitter receives in-phase and quadrature components (real and imaginary components) of the master signal (top two figures).  It then obtains an initial phase estimate (middle figure) from these samples. The secondary tracks the phase drift of the subcarriers beginning at the 10th symbol (second from bottom figure) and uses a filter to predict its value a few symbols later (bottom figure).}}
 \label{phase_sync}
\end{figure*}

Our system setup is presented in Figure \ref{testbed}. It consists of
a primary transmitter, three secondary transmitters and four receivers. The
main sender uses a single RF front-end configured in transmit mode,
placing an 18 MHz shaping filter around the transmitted signal. The
secondary senders use an RF front-end in receive mode and a second RF
front-end in transmit mode, with a 12 MHz shaping filter. As
mentioned previously, the pilots used in phase tracking are outside
the secondary's transmission band, therefore the secondary transmitter
will not interfere with the pilot signals from the main transmitter.
The series of experiments is intended to test the accuracy of the
synchronization, the efficiency of channel separation and the extent
to which we achieve the theoretical gains that multiuser MIMO promises
in our setup.
 
\subsection{Synchronization Accuracy}\label{sync_accur} 

In this particular experiment we have placed two transmitters and
two receivers at random locations. We placed a third RF front-end
on the secondary sender and configured it in receive mode. The
secondary transmitter samples its own synthesized signal over a wired
feedback loop and compares it with the main transmitter's signal. The
synchronization circuit measures and records the phase differences
between these two signals. Since we use the primary transmission as a
reference, in this experiment we do not broadcast the signal
synthesized by the secondary transmitter in order to protect the
primary transmission from unintended interference. 

We have modified the synchronization circuit to produce a signal that
is not only phase synchronous with that of the primary transmitter
but has the exact same phase when observed from the secondary
transmitter. To achieve this, the circuit estimates the phase rotation
that is induced between the DAC of the secondary transmitter and the
ADC through which the synthesized signal is resampled. It then
compensates for this rotation by subtracting this value from the
initial phase estimate. It is worth noting that this rotation
corresponds to the propagation delay through the feedback circuit and
is constant for different packet transmissions, as determined through
measurements. The result was a synthesized signal that closely follows
the phase of the signal broadcast by the master transmitter, as
illustrated in Figure \ref{phase_sync}. The figure illustrates the
initial phase acquisition process, the initial phase estimation, the
tracking and estimation of the phase drift, as well as the
synthesis of the new signal. The phase discontinuities appearing in
the main transmitter's signal are due to the presence of the PN sequence
along with a temporary disturbance needed in order to tune the
feedback circuit.

\begin{figure}[htb]
 \centering
 \includegraphics[width=.4\textwidth]{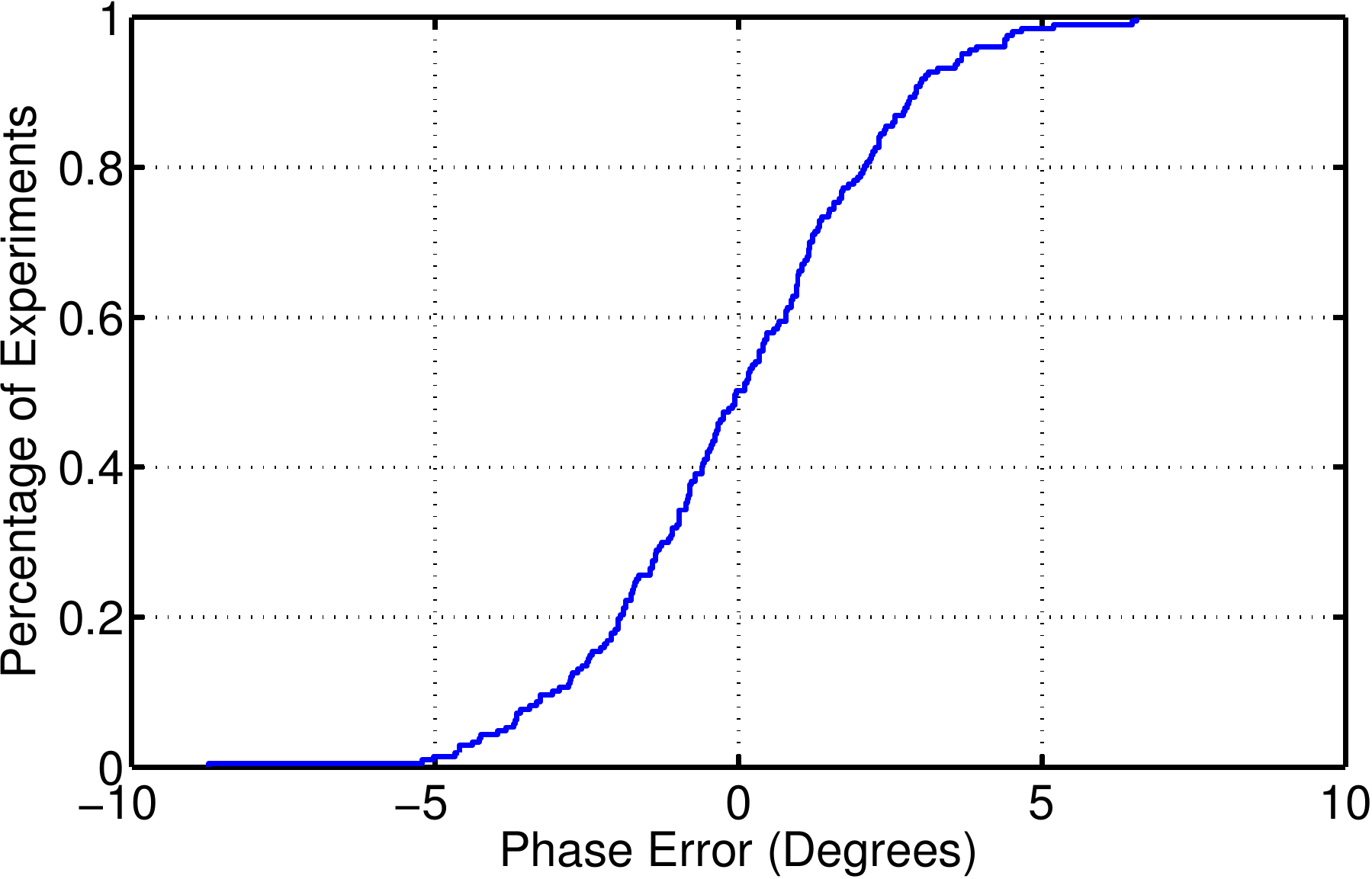}
 \caption{The Precision of the Phase Synchronization. \textnormal{AirSync achieves phase synchronization within a few degrees of the source signal.}}
 \label{cdf}
\end{figure}

Figure \ref{cdf} illustrates the CDF of the synchronization
error between the secondary transmitter and the primary
transmitter. The error is measured on a frame-to-frame basis using the
feedback circuit. In decimal degree values, 
the standard deviation is 2.37 degrees. The 95th percentile of the
synchronization error is at most 4.5 degrees.

The radios were placed in a typical office environment. We have
measured the SNR value of the synchronization pilots in the signal
received by the secondary transmitter to be around 28.5 dB above the
noise floor. This is easily achievable between typically placed access
points.

\subsection{Beamforming gain}

Our second experiment was done using two transmitters and a receiver
with the secondary transmitter broadcasting a secondary signal over
the air. We measured the channel coefficients between the two
transmitters and the receivers using standard downlink channel
estimation techniques and arranged the amplitudes and the phases of
the transmitted signals such that at one of the receivers the
amplitudes of the two transmitted signals would be equal while the
phases would align. The maximal theoretic power gain over transmitting
the two signals independently is 3.01dB. We compared the average power of the
individual transmissions from the two senders to the average power of a
beamformed joint transmission. Our measurements show an average
gain of 2.98 dB, which is consistent with the precision of the
synchronization determined in the previous experiment.

This result shows that for all practical purposes we are able to
achieve the full beamforming gain in our testbed.

\subsection{Zero-Forcing Accuracy}

\begin{figure}[h]
 \centering
 \includegraphics[width=.4\textwidth]{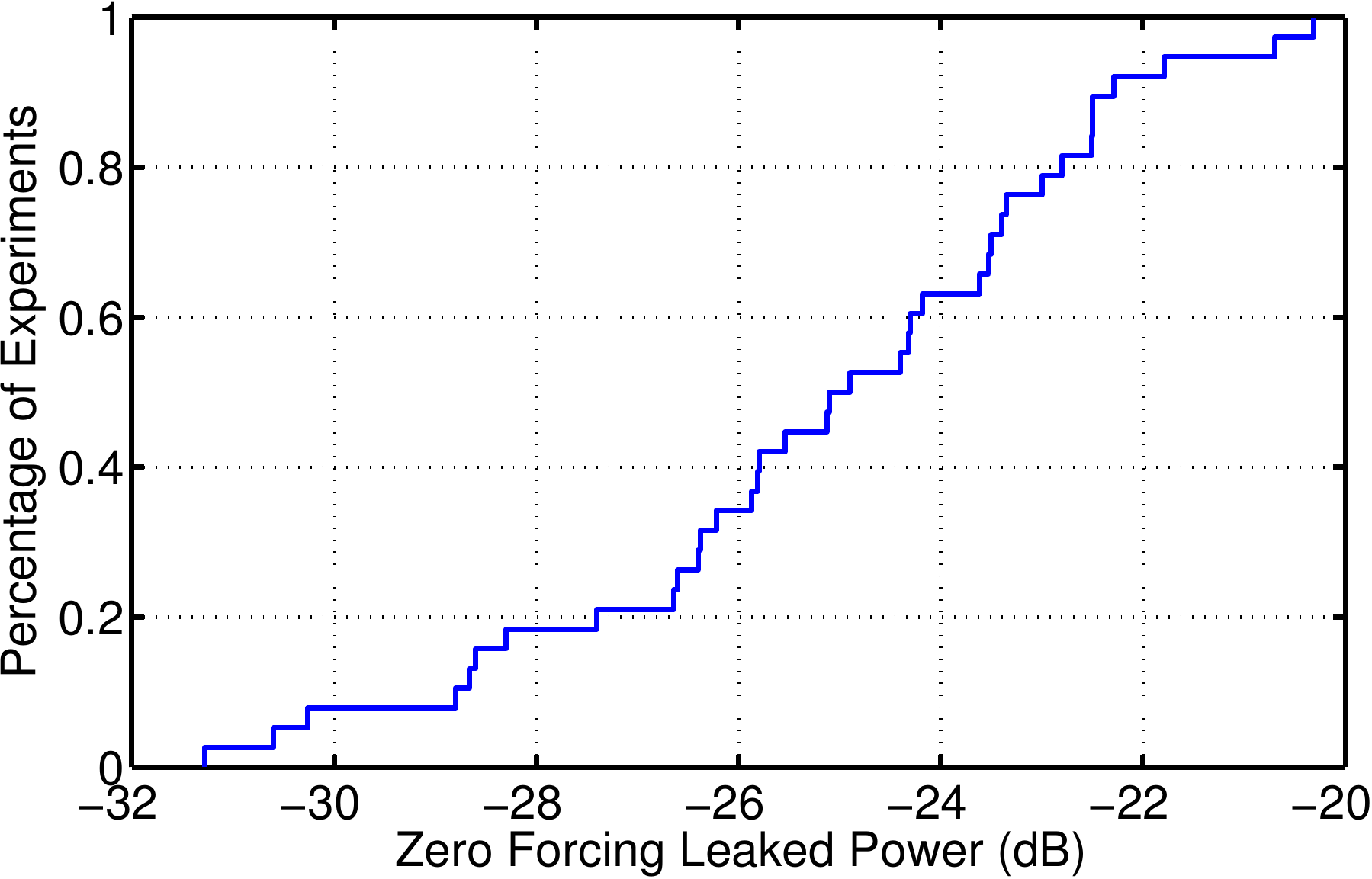}
 \caption{The Power Leakage of Zero-Forcing. \textnormal{The leaked power is significantly smaller than the total transmitted power, transforming each receiver's channel into a high SINR channel.}}
 \label{zfcdf}
\end{figure}

The following experiment measures the amount of power which is
inadvertently leaked when using Zero-Forcing to non-targeted receivers
due to synchronization errors. Again we have placed two transmitters
and a receiver at random locations in our testbed. We have estimated
the channel coefficients and arranged for two equal amplitude tones
from the two transmitters to sum as closely as possible to zero. The
residual power is the leaked power due to angle mismatching. Figure
\ref{zfcdf} illustrates the CDF of this residual power for different
measurements. The average power leaked is -24.46 dB of the total
transmitted power.
This establishes that Zero-Forcing is capable of almost completely
eliminating interference at non-targeted receiver locations.

\subsection{Zero-Forcing Beamforming Data Transmission}

\begin{figure}[h]
 \centering
 \includegraphics[width=.4\textwidth]{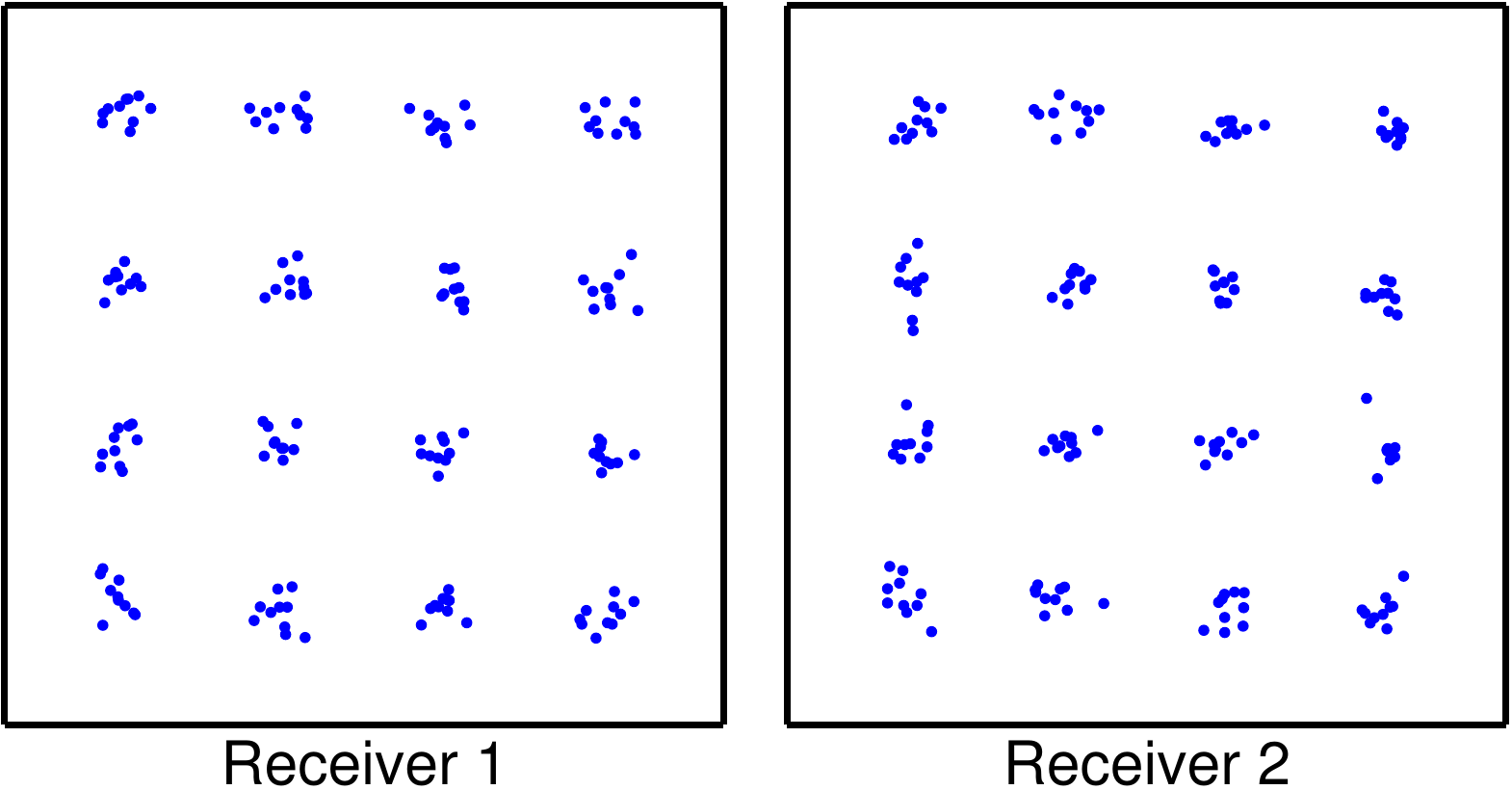}
 \caption{\textnormal{The scattering diagram for two independent data streams transmitted concurrently using ZFBF demonstrates that AirSync achieves complete separation of the user channels.}}
 \label{scattering}
\end{figure}

This experiment transmits data from two transmitters to two receivers
using ZFBF. We have used symbols chosen independently from a QAM-16
constellation at similar power levels. The scattering plots in Figure
\ref{scattering} illustrate the received signals at the two receivers.
From the figure it is clear that we have created two separate
channels, achieving thus a multiplexing factor of two over
point-to-point transmissions. The actual rates achieved will depend on
the quality of the two channels.

We would like to compare the performance of the multiuser MIMO system
to a current standard. In current enterprise WiFi networks
transmissions within a small area occur from single access points to
single clients and are separated in time using TDMA. We use the best
achievable point-to-point rate as an upper limit for the rates that
the TDMA approach can achieve and compare the rates achieved by
our system.

The SINR values at the two receivers are 29 dB and 26 dB respectively.
In the same experiment, we measured the best point-to-point link to
have a 32 dB SNR value. Using Shannon's formula, these values
translate to maximally achievable rates of 9.96 bits/second/Hz
(bps/Hz)for the point-to-point channel and 18.27 bps/Hz for the
compound MIMO channel. Thus, when using ideal codes, we achieve a
multiplexing rate gain of 1.83, which is close to the theoretical
value of 2.

At all the mentioned SNR levels 802.11g (a point-to-point standard)
uses the same 64-QAM modulation, resulting in a rate
of 6 bps/Hz (ignoring the error correcting code overhead, which is
identical for all three SNR levels). Thus, we can say that both
of the channels obtained through zero-forcing support WiFi operation
at the highest commonly used rates and therefore equivalate to
independent WiFi channels. We conclude that, using practical
modulations, the experimental multiplexing gain equals the theoretical
value of 2.

\subsection{Tomlinson-Harashima precoding}

\begin{figure*}[tb]
 \centering
 \includegraphics[width=.95\textwidth]{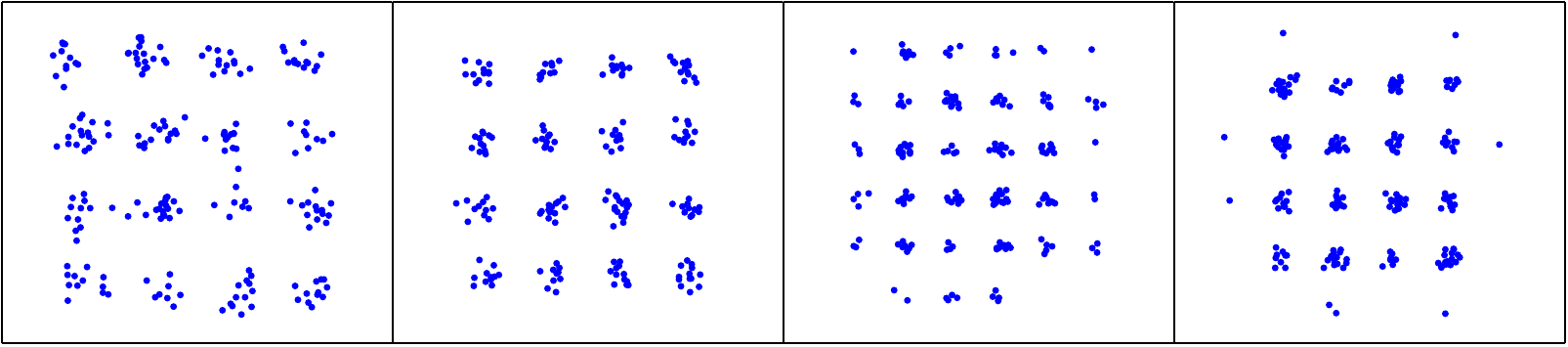}
 \caption{Tomlinson-Harashima precoding based on on QAM-16
   constellations. The achieved spectral efficiency is 16
   bits/second/Hz}
 \label{thnc}
\end{figure*}

The final experiment uses four transmitters and four receivers.  We
employ Tomlinson-Harashima precoding.  The results are illustrated in
Figure \ref{thnc}, which presents the four distinct wireless channels
created for the four users. Thus, we have achieved a multiplexing
factor of 4. As before, the actual rate gains will depend on the
quality of the channels.

We measured the SINR values of the four channels to be 16.8 dB, 19.2
dB, 21.4 dB and 20.8 dB. The lower SINR values are caused by increased
levels of power leakage due to the presence of more transmissions to
other receivers (see Figure \ref{zfcdf} for the distribution of leaked
power from a single interfering transmission). Again, the Shannon rate
formula predicts achievable channel rates of 5.6 bps/Hz, 6.4 bps/Hz,
7.11 bps/Hz and 6.91 bps/Hz. The sum rate is 26 bps/Hz. As mentioned
before, the best point-to-point channel in our setup has a quality
level of 32 dB, allowing for 9.96 bps/Hz. Therefore the rate
gain is about 2.6 when using four degrees of freedom and ideal codes.

More practically, we can compare the performance of our system when
employing an extended 16-QAM constellation on every channel with the
performance of 802.11g using a typical modulation. At 32 dB SNR,
802.11g would use a 64-QAM constellation and achieve (ignoring the
error correcting code overhead) a spectral efficiency of 6 bps/Hz. In
the MIMO case, we can achieve a sum rate of 16 bps/Hz using four
16-QAM constellations, leading to a multiplexing gain of 2.66 under
practical modulations, while the theoretical value is 4. In a
commercial implementation, we expect the leakage to be further reduced
and we expect to be able to come closer of a rate gain of 4. In
general, nearing the theoretical rate gains through spatial
multiplexing requires precise channel state information and tight
synchronization, as evidenced by our experiments.

%% file: medium.tex
\section{Medium Access Control}
\label{section:medium}

Given that we have achieved the necessary synchronization accuracy between access points and realized the full multiplexing gain,
we turn to the large body of work on optimal scheduling for centralized multiuser MIMO systems (see for example 
\cite{caire_isit,dinic_sidiropoulos}). 
Inspired by this work, we propose 
a MAC layer that significantly departs from the classic networking layered architectural model and adopts a cross-layer ``PHY/MAC'' design strategy. 

\subsection{High level description}

{\bf Time Division Duplexing.}  First, we consider the issue of
allocating air time and frequency spectrum between the uplink and the
downlink.  We can choose between two natural strategies for separating
the uplink from the downlink: time division duplex (TDD) and frequency
division duplex (FDD). TDD has the following two advantages. First,
with TDD one can exploit channel reciprocity and measure the uplink
channel, using pilots from the users to infer the downlink channel.
 
In the case of FDD, an explicit closed-loop channel
estimation (from the downlink pilots sent by the access points) and
feedback (from the clients to the server) needs to be implemented,
with a protocol overhead that increases linearly with the number of
jointly precoded access point antennas
\cite{JoseChannelEstimation}. Second, TDD is ideally suited for the
transport of asymmetric traffic, as is typical in an enterprise WiFi
scenario, whereas an FDD system provides less flexibility for managing
different traffic patterns. Specifically, with TDD, the downlink
channel estimation procedure and the downlink time reservation
proposed in the 802.11ac standard \cite{802.11ac} can be applied to our
distributed MIMO system as well.

We shall
consider the scheduling of users in the uplink and downlink periods
separately. In the uplink, clients compete for bandwidth using regular
CSMA/CA. Thus, in the rest of this section we focus on the downlink. We note 
here that in order for our system to be backward compatible with legacy 802.11 clients
and access points, protection mechanisms and modes of operation have to be 
implemented. Such mechanisms are described in the 802.11n/ac standards \cite{802.11ac} 
where, using RTS/CTS, CTS-to-self frames and legacy format preambles,
nearby devices can sense that the channel is in use and avoid collisions.

{\bf Downlink scheduling.}  The central server keeps track of packet
queue sizes and other readily available QoS information, e.g. the time
since these queues have been served last.  It then selects a subset of
users to transmit to at each downlink time slot.
In the following we discuss in detail how the server selects these
users at each time slot when ZFBF is the precoding scheme of choice.
A similar approach can be applied to THP precoding with minimal changes.

The user selection and power allocation problem for linear Zero-Forcing
precoding has a rich literature (for example
\cite{dinic_sidiropoulos,yoo-goldsmith,fuchs07}). 
Conceptually, this optimization problem can be solved
by exhaustively searching over all feasible subsets of users,
optimizing a weighted rate function under some general power
constraints. In practice, greedy algorithms have proven to provide
excellent results at moderate complexity \cite{caire_isit,
  dinic_sidiropoulos}.

\begin{figure}[t]
  \centering
  \includegraphics[width=.4\textwidth]{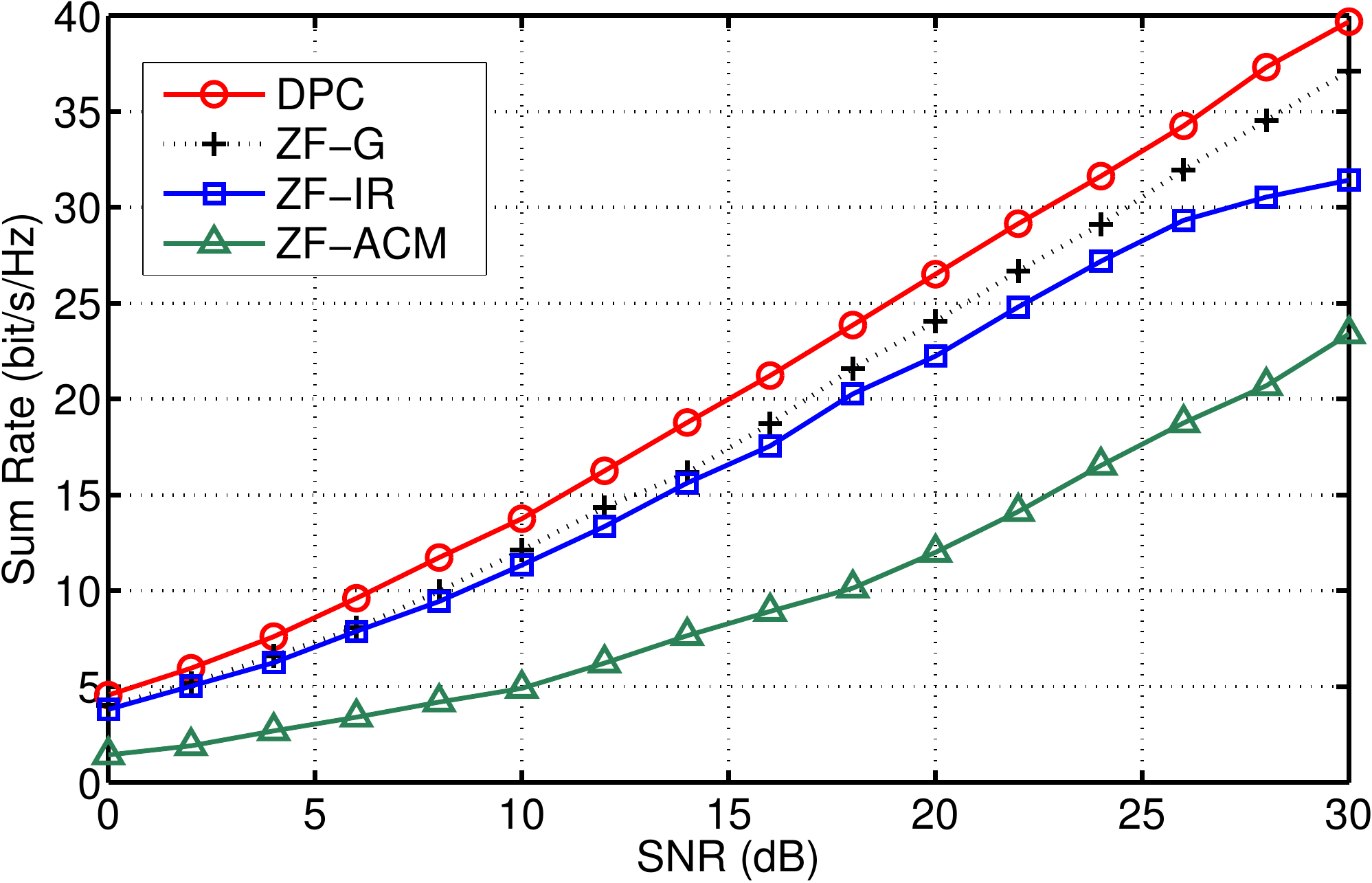}
  \caption{Comparison of the sum-rates for various greedy zero-forcing schemes. \textnormal{.}}
\label{sumratecomp}
 \end{figure}

We begin by evaluating the achievable sum-rate using such a
greedy policy. Firstly, the use of coding rates equal to the
corresponding Gaussian channel capacity $\log(1 + \text{SINR})$ is
overly idealized; by mapping the SINRs into a discrete set of
modulation and coding schemes (MCSs), we can model a more realistic
scenario.  For the sake of simplicity we assume that we can choose the
best scheme based on the received SINR.  Table \ref{tbl:acm} provides
one such mapping that corresponds to the 9 mandatory MCSs of 802.11ac
\cite{802.11ac}, keeping in mind that mappings vary by vendor or may
be dynamically chosen in practical scenarios. In Figure
\ref{sumratecomp} the sum rates for the two schemes, greedy ZFBF with
ideal rates (ZF-G) and the adaptive coding and modulation (ACM)
scenario (ZF-ACM) described above are evaluated for multiple SNRs in
the case of 10 clients and 4 total access points antennas. For
purposes of reference, the optimal, capacity-achieving Dirty Paper
Coding (DPC) \cite{costa-dpc} precoding technique is shown in the same
plot.

The huge gap between ZF-ACM and the ideal ZF motivates us to turn our attention to more flexible ways of allocating the rates in the multiuser MIMO scenario. The current standard, 802.11n, offers many code
combinations to fully utilize the capacity of the MIMO channel. Since
a multiuser MIMO system serves multiple users in the same time slot,
an even larger set of rates and codes would have to be supported for
efficiently using capacity. An attractive and innovative approach is the use
of {\em rateless codes} (e.g., Raptor codes \cite{raptor,
 raptor_BIMSC} and the recently proposed Spinal codes \cite{spinal})
at the physical layer, in a so-called {\em Incremental Redundancy}
(IR) configuration (see
\cite{caire-tuninetti,scv04,soljanin-incremental}), as already
exemplified by Strider, to decrease the signaling and
retransmission overhead. In an ideal rateless coding adaptation scenario, we would achieve the {\em coded modulation capacity} of a fixed large QAM constellation. In Figure \ref{sumratecomp} the performance of greedy zero-forcing with such an ideal rateless code (ZF-IR) is also depicted for an ideal family of random rateless codes based on a 256-QAM constellation. It is immediately obvious that the gains of using this IR configuration are tremendous in comparison with classic ACM.

\begin{table}[h]
  \begin{center}
    \begin{tabular}{| c | c | c | c |}
    \hline
    802.11ac MCS Index & Modulation & Code Rate & SNR Range \\ \hline
    0 & BPSK & 1/2 & $\geq$ 2dB \\ \hline
    1 & QPSK & 1/2 & $\geq$ 5dB \\ \hline
    2 & QPSK & 3/4 & $\geq$ 8dB \\ \hline
    3 & 16-QAM & 1/2 & $\geq$ 12dB \\ \hline
    4 & 16-QAM & 3/4 & $\geq$ 15dB \\ \hline
    5 & 64-QAM & 2/3 & $\geq$ 18dB \\ \hline
    6 & 64-QAM & 3/4 & $\geq$ 21dB \\ \hline
    7 & 64-QAM & 5/6 & $\geq$ 24dB \\ \hline
	8 & 256-QAM & 3/4 & $\geq$ 27dB \\
    \hline
    \end{tabular}
  \end{center}
  \caption{Modulation/Coding pairs from IEEE 802.11ac and the corresponding SNRs at which they can be selected}
  \label{tbl:acm}
\end{table}

\subsection{Protocol Design}

Our protocol design focuses on the downlink channel. 
The MAC layer protocol is tuned for enabling
multiuser MIMO broadcasts. 
The crucial design constraint is to provide
the central server with timely estimates of the channel state
information for all clients to which it is about to transmit or which
are considered for the next round of transmissions.  For this purpose,
we can collect channel estimates either at the access points through
uplink pilots (based on TDD reciprocity) or at the receivers using a
standard downlink estimation procedure as described in
802.11ac.
The central server uses the estimates to select a set of clients for
the following transmission slots, according to the scheduling
algorithms introduced earlier.

The choice between uplink and downlink estimation has an important
impact on the design of the synchronization system. When using
downlink pilots, the system must guarantee that the effective channel
matrix before the receiver, that is the product of the two right-most
matrices in (\ref{eq:effective}), is constant between packet
transmissions. This calls for actively aligning the phases of the
transmitters (the right-most matrix) for every packet. In contrast,
when using uplink estimation, the phase shifts induced by frame
misalignment on the uplink and the downlink path cancel each other,
allowing the access points to skip the phase alignment step. 

Our protocol design follows the lines of 802.11ac \cite{802.11ac}:
before a downlink transmission period the access points broadcast a
request for a number of clients to estimate their channels based on a
channel probing message broadcasted shortly after. The access points
then transmit requests for feedback in succession to each targeted
client and wait for the corresponding feedback. Once all the
information has been collected, the downlink period can begin.  We
note that the use of a STBC for control frames can improve their
robustness, given that from a client perspective the phases of the
access points are essentially random during this phase.

The downlink data packet starts with a transmission from the main
sender containing a pseudo-noise sequence used to achieve frame
alignment by the transmitters and for block boundary detection by the
receivers. The master access point then transmits the first set of
channel estimation pilots which are used by the other access points to
determine the initial phases of the subcarrier tones, as described in
Section \ref{sec:hardware}.  After this point, all access points take
part in the downlink transmission.  

We tested each component of the downlink and uplink protocol
slots. However, since our radios do not switch from receive to
transmit in a timely manner, we could not perform complete real-time
MAC experiments.

{\bf Overhead.}
A note on the overhead of the above MAC is in order.
the overhead of our MAC is not more than that of 802.11n.
The additional signaling overhead comes from requiring a few frames to predict the 
initial phase, and a few frames to dictate the MAC addresses of the nodes from which 
we wish to request channel state information for the next time slot. Even with very conservative 
estimates this will be less than a 20\% increase in header time duration over that of a traditional 802.11 system. 
Note, however, that we get a bandwidth increase that grows almost linearly in the number of clients. This 
means that our overhead, normalized such that we consider the total control bits over the total data bits 
transmitted during a fixed airtime slot, is much less than in a traditional 802.11 system.

%% file: discussion.tex
\section{Discussion and future work}
\label{sec:discussion}

Our future work is concerned with improving the robustness of the
parameter estimators used in the synchronization system while reducing
the synchronization overhead. We plan to complete a MAC layer
implementation which relies on uplink pilots estimates (based on
channel reciprocity) for obtaining low-overhead estimates of the
channel matrix. Another research topic is making Airsync scalable
through semi-decentralized precoding and the use of a hierarchical
structure. Finally, we would like to extend our system by implementing
a joint PHY/MAC layer based on an actual family of rateless codes and
an incremental redundancy-based MAC layer.

%% file: ton_airsync.bbl
\begin{thebibliography}{10}

\bibitem{cisco_report}
{Cisco Visual Networking Index: Global Mobile Data Traffic Forecast Update,
  2011 - 2016}.
\newblock Technical report, {Cisco VNI}, Feb 2012.

\bibitem{802.11ac}
{IEEE Draft Standard for IT - Telecommunications and Information Exchange
  Between Systems - LAN/MAN - Specific Requirements - Part 11: Wireless LAN
  Medium Access Control and Physical Layer Specifications - Amd 4: Enhancements
  for Very High Throughput for operation in bands below 6GHz}.
\newblock {\em IEEE P802.11ac/D3.0, June 2012}, pages 1 --385, 11 2012.

\bibitem{aryafar10design}
E.~Aryafar, N.~Anand, T.~Salonidis, and E.~W. Knightly.
\newblock Design and experimental evaluation of multi-user beamforming in
  wireless {LAN}s.
\newblock In {\em {ACM} Mobi{C}om}, Chicago, IL, 2010.

\bibitem{tosato-boccardi-caire}
F.~Boccardi, F.~Tosato, and G.~Caire.
\newblock Precoding schemes for the mimo-gbc.
\newblock In {\em Communications, 2006 International Zurich Seminar on}, pages
  10 --13, 0-0 2006.

\bibitem{caire2010multiuser}
G.~Caire, N.~Jindal, M.~Kobayashi, and N.~Ravindran.
\newblock Multiuser {MIMO} achievable rates with downlink training and channel
  state feedback.
\newblock {\em {IEEE} Trans. Inf. Theory}, 56(6):2845--2866, 2010.

\bibitem{caire-shamai}
G.~Caire and S.~Shamai.
\newblock On the achievable throughput of a multiantenna gaussian broadcast
  channel.
\newblock {\em {IEEE} Trans. Inf. Theory}, 49(7):1691 -- 1706, Jul. 2003.

\bibitem{caire-tuninetti}
G.~Caire and D.~Tuninetti.
\newblock The throughput of hybrid-arq protocols for the gaussian collision
  channel.
\newblock {\em {IEEE} Trans. Inf. Theory}, 47(5):1971 --1988, Jul. 2001.

\bibitem{chintalapudi12wifi}
K.~Chintalapudi, B.~Radunovic, V.~Balan, M.~Buettener, S.~Yerramalli, V.~Navda,
  and R.~Ramjee.
\newblock Wifi-nc: Wifi over narrow channels.
\newblock In {\em Proceedings of the 9th USENIX conference on Networked Systems
  Design and Implementation}, NSDI'12, pages 4--4, Berkeley, CA, USA, 2012.
  USENIX Association.

\bibitem{katti}
J.~I. Choi, M.~Jain, K.~Srinivasan, P.~Levis, and S.~Katti.
\newblock Achieving single channel, full duplex wireless communication.
\newblock In {\em IEEE MobiCom}, Chicago, IL, 2010.

\bibitem{costa-dpc}
M.~Costa.
\newblock Writing on dirty paper (corresp.).
\newblock {\em {IEEE} Trans. Inf. Theory}, 29(3):439--441, May 1983.

\bibitem{coverthomas}
T.~M. Cover and J.~A. Thomas.
\newblock {\em Elements of information theory}.
\newblock Wiley-Interscience, New York, NY, USA, 1991.

\bibitem{dinic_sidiropoulos}
G.~Dimic and N.~Sidiropoulos.
\newblock On downlink beamforming with greedy user selection: performance
  analysis and a simple new algorithm.
\newblock {\em {IEEE} Trans. Signal Process.}, 53(10):3857 -- 3868, Oct. 2005.

\bibitem{ashutosh}
M.~Duarte, C.~Dick, and A.~Sabharwal.
\newblock Experiment-driven characterization of full-duplex wireless systems.
\newblock {\em CoRR}, abs/1107.1276, 2011.

\bibitem{erez-zamir-shamai}
U.~Erez, S.~Shamai, and R.~Zamir.
\newblock Capacity and lattice strategies for canceling known interference.
\newblock {\em Information Theory, IEEE Transactions on}, 51(11):3820 -- 3833,
  nov. 2005.

\bibitem{raptor_BIMSC}
O.~Etesami and A.~Shokrollahi.
\newblock Raptor codes on binary memoryless symmetric channels.
\newblock {\em {IEEE} Trans. Inf. Theory}, 52(5):2033 -- 2051, May 2006.

\bibitem{forney-eyuboglu}
J.~Forney, G.D. and M.~Eyuboglu.
\newblock Combined equalization and coding using precoding.
\newblock {\em Communications Magazine, IEEE}, 29(12):25 --34, dec. 1991.

\bibitem{foschini}
G.~Foschini and M.~Gans.
\newblock On limits of wireless communications in a fading environment when
  using multiple antennas.
\newblock {\em Wireless Personal Communications}, 6:311--335, 1998.

\bibitem{fuchs07}
M.~Fuchs, G.~Del~Galdo, and M.~Haardt.
\newblock Low-complexity space-time-frequency scheduling for mimo systems with
  sdma.
\newblock {\em Vehicular Technology, IEEE Transactions on}, 56(5):2775 --2784,
  sept. 2007.

\bibitem{golakotta08zigzag}
S.~Gollakota and D.~Katabi.
\newblock Zigzag decoding: combating hidden terminals in wireless networks.
\newblock In {\em ACM SIGCOMM}, Seattle, WA, 2008.

\bibitem{gollakota09interference}
S.~Gollakota, S.~D. Perli, and D.~Katabi.
\newblock Interference alignment and cancellation.
\newblock In {\em ACM SIGCOMM}, Barcelona, Spain, 2009.

\bibitem{JoseChannelEstimation}
J.~Jose, A.~Ashikhmin, P.~Whiting, and S.~Vishwanath.
\newblock Channel estimation and linear precoding in multiuser multiple-antenna
  {TDD} systems.
\newblock {\em {IEEE} Trans. Veh. Technol.}, 60(5):2102 --2116, Jun. 2011.

\bibitem{jose2008scheduling}
J.~Jose, A.~Ashikhmint, P.~Whiting, and S.~Vishwanath.
\newblock Scheduling and pre-conditioning in multi-user {MIMO TDD} systems.
\newblock In {\em IEEE ICC}, Beijing, China, 2008.

\bibitem{caire_isit}
M.~Kobayashi and G.~Caire.
\newblock Joint beamforming and scheduling for a {MIMO} downlink with random
  arrivals.
\newblock In {\em IEEE ISIT}, Seattle, WA, Jul. 2006.

\bibitem{lapidoth_allerton}
A.~Lapidoth, S.~Shamai, and M.~A. Wigger.
\newblock On the capacity of fading {MIMO} broadcast channels with imperfect
  transmitter side-information.
\newblock In {\em 43rd Annual Allerton Conference}, Monticello, IL, 2005.

\bibitem{soljanin-incremental}
C.~Lott, O.~Milenkovic, and E.~Soljanin.
\newblock Hybrid {ARQ}: Theory, state of the art and future directions.
\newblock In {\em IEEE ITW on Information Theory for Wireless Networks},
  Solstrand, Norway, Jul. 2007.

\bibitem{molischbook}
A.~Molisch.
\newblock {\em Wireless Communications}.
\newblock Wiley-IEEE Press, 2005.

\bibitem{spinal}
J.~Perry, H.~Balakrishnan, and D.~Shah.
\newblock Rateless spinal codes.
\newblock In {\em ACM HotNets}, Cambridge, Massachusetts, 2011.

\bibitem{proakis}
J.~Proakis and M.~Salehi.
\newblock {\em Digital communications}.
\newblock McGraw-Hill, New York, NY, 2007.

\bibitem{rahul10sourcesync}
H.~Rahul, H.~Hassanieh, and D.~Katabi.
\newblock {S}ource{S}ync: a distributed wireless architecture for exploiting
  sender diversity.
\newblock In {\em ACM SIGCOMM}, New Delhi, India, 2010.

\bibitem{warp}
{Rice University}.
\newblock Rice university warp project.

\bibitem{scv04}
S.~Sesia, G.~Caire, and G.~Vivier.
\newblock Incremental redundancy hybrid {ARQ} schemes based on low-density
  parity-check codes.
\newblock {\em {IEEE} Trans. Commun.}, 52(8):1311 -- 1321, Aug. 2004.

\bibitem{argos}
C.~Shepard, H.~Yu, N.~Anand, L.~Li, T.~Marzetta, R.~Yang, and L.~Zhong.
\newblock Argos: Practical many-antenna base stations.
\newblock {\em SIGCOMM Comput. Commun. Rev.}, Aug. 2012.

\bibitem{raptor}
A.~Shokrollahi.
\newblock Raptor codes.
\newblock {\em {IEEE} Trans. Inf. Theory}, 52(6):2551 --2567, Jun. 2006.

\bibitem{spencer2004introduction}
Q.~Spencer, C.~Peel, A.~Swindlehurst, and M.~Haardt.
\newblock An introduction to the multi-user {MIMO} downlink.
\newblock {\em {IEEE} Commun. Mag.}, 2004.

\bibitem{tan10fine}
K.~Tan, J.~Fang, Y.~Zhang, S.~Chen, L.~Shi, J.~Zhang, and Y.~Zhang.
\newblock Fine-grained channel access in wireless {LAN}.
\newblock In {\em ACM SIGCOMM}, New Delhi, India, 2010.

\bibitem{telatar_mimo_cap}
E.~Telatar.
\newblock Capacity of multi-antenna gaussian channels.
\newblock {\em European Transactions on Telecommunications}, 10(6):585--595,
  1999.

\bibitem{wc_book_tse}
D.~Tse and P.~Viswanath.
\newblock {\em Fundamentals of Wireless Communication}.
\newblock Cambridge University Press, New York, NY, 2005.

\bibitem{vaze_degrees}
C.~S. Vaze and M.~K. Varanasi.
\newblock The degrees of freedom regions of {MIMO} broadcast, interference, and
  cognitive radio channels with no {CSIT}.
\newblock {\em CoRR}, abs/0909.5424, 2009.

\bibitem{verdu_mud}
S.~Verdu.
\newblock {\em Multiuser Detection}.
\newblock Cambridge University Press, New York, NY, 1998.

\bibitem{weingarten}
H.~Weingarten, Y.~Steinberg, and S.~Shamai.
\newblock The capacity region of the gaussian multiple-input multiple-output
  broadcast channel.
\newblock {\em {IEEE} Trans. Inf. Theory}, 52(9):3936 --3964, Sept. 2006.

\bibitem{windpassinger-huber}
C.~Windpassinger, R.~Fischer, T.~Vencel, and J.~Huber.
\newblock Precoding in multiantenna and multiuser communications.
\newblock {\em Wireless Communications, IEEE Transactions on}, 3(4):1305 --
  1316, Jul. 2004.

\bibitem{yoo-goldsmith}
T.~Yoo and A.~Goldsmith.
\newblock On the optimality of multiantenna broadcast scheduling using
  zero-forcing beamforming.
\newblock {\em {IEEE} J. Sel. Areas Commun.}, 24(3):528 -- 541, Mar. 2006.

\end{thebibliography}
